\documentclass[preprint,5p, a4paper, sort&compress,authoryear,hyperref]{elsarticle}

\usepackage[T1]{fontenc}
\usepackage[fleqn]{amsmath}
\usepackage{amssymb, amsfonts}
\usepackage{booktabs} 
\usepackage{mathtools}
\usepackage{commath}
\usepackage{microtype}

\usepackage{amsthm}
\usepackage{mathdots}
\usepackage{siunitx}

\allowdisplaybreaks

\newcommand{\myreferences}{new_extracted}

\usepackage{tikz}
\usepackage{pgfplots}
\pgfplotsset{compat=newest}
\usetikzlibrary{shapes,fit,arrows,decorations.markings,matrix,plotmarks,positioning,fit,spy,patterns,shadows,calc,backgrounds}
\usepackage{subfig}

\usepackage{graphicx}
\usepackage[abs]{overpic}

\usepackage[applemac]{inputenc}
\usepackage[nolist]{acronym}

\usepackage{enumerate}
\newcounter{subeqn} %

\newtheorem{theorem}{Theorem}[section]
\newtheorem{corollary}{Corollary}[theorem]
\newtheorem{lemma}{Lemma}[section]
\newtheorem{proposition}{Proposition}[section]
\newtheorem{assumption}{Assumption}
\newtheorem{definition}{Definition}%[section]
\newtheorem{alg}{Algorithm}%[section]
\newtheorem{example}{Example}%[section]
\newtheorem{remark}{Remark}

\newcommand{\ts}[1]{{\textnormal{#1}}}
\newcommand{\ie}{\emph{i.e.},\ }
\newcommand{\eg}{\emph{e.g.}}

\newcommand{\Rset}{\mathbb{R}}
\newcommand{\mc}{\mathcal}
\newcommand{\mb}{\mathbf}
\newcommand{\mbb}{\mathbb}
\newcommand{\mf}{\mathfrak}

\DeclareMathAlphabet\mathbfcal{OMS}{cmsy}{b}{n}

\DeclarePairedDelimiter{\diagpars}{(}{)}
\newcommand{\diag}{\operatorname{diag}\diagpars}
\DeclarePairedDelimiter{\intpars}{(}{)}
\newcommand{\interior}{\operatorname{interior}\intpars}

\newcommand\inN[2]{#1\in\mathcal{M}_{#2}}

\newcommand\st{:}

\newlength{\hatchspread}
\newlength{\hatchthickness}
\newlength{\hatchshift}
\newcommand{\hatchcolor}{}
\tikzset{hatchspread/.code={\setlength{\hatchspread}{#1}},
         hatchthickness/.code={\setlength{\hatchthickness}{#1}},
         hatchshift/.code={\setlength{\hatchshift}{#1}},% must be >= 0
         hatchcolor/.code={\renewcommand{\hatchcolor}{#1}}}
\tikzset{hatchspread=3pt,
         hatchthickness=0.4pt,
         hatchshift=0pt,% must be >= 0
         hatchcolor=black}
\pgfdeclarepatternformonly[\hatchspread,\hatchthickness,\hatchshift,\hatchcolor]% variables
   {nwl}% name
   {\pgfqpoint{\dimexpr-2\hatchthickness}{\dimexpr-2\hatchthickness}}% lower left corner
   {\pgfqpoint{\dimexpr\hatchspread+2\hatchthickness}{\dimexpr\hatchspread+2\hatchthickness}}% upper right corner
   {\pgfqpoint{\dimexpr\hatchspread}{\dimexpr\hatchspread}}% tile size
   {% shape description
    \pgfsetlinewidth{\hatchthickness}
    \pgfpathmoveto{\pgfqpoint{0pt}{\dimexpr\hatchspread+\hatchshift}}
    \pgfpathlineto{\pgfqpoint{\dimexpr\hatchspread+0.15pt+\hatchshift}{-0.15pt}}
    \ifdim \hatchshift > 0pt
      \pgfpathmoveto{\pgfqpoint{0pt}{\hatchshift}}
      \pgfpathlineto{\pgfqpoint{\dimexpr0.15pt+\hatchshift}{-0.15pt}}
    \fi
    \pgfsetstrokecolor{\hatchcolor}
    \pgfusepath{stroke}
   }
\pgfdeclarepatternformonly[\hatchspread,\hatchthickness,\hatchshift,\hatchcolor]% variables
   {nwld}% name
   {\pgfqpoint{\dimexpr-2\hatchthickness}{\dimexpr-2\hatchthickness}}% lower left corner
   {\pgfqpoint{\dimexpr\hatchspread+2\hatchthickness}{\dimexpr\hatchspread+2\hatchthickness}}% upper right corner
   {\pgfqpoint{\dimexpr\hatchspread}{\dimexpr\hatchspread}}% tile size
   {% shape description
    \pgfsetlinewidth{\hatchthickness}
    \pgfpathmoveto{\pgfqpoint{0pt}{\dimexpr\hatchspread+\hatchshift}}
    \pgfpathlineto{\pgfqpoint{\dimexpr\hatchspread+0.15pt+\hatchshift}{-0.15pt}}
    \ifdim \hatchshift > 0pt
      \pgfpathmoveto{\pgfqpoint{0pt}{\hatchshift}}
      \pgfpathlineto{\pgfqpoint{\dimexpr0.15pt+\hatchshift}{-0.15pt}}
    \fi
    \pgfsetstrokecolor{\hatchcolor}
    \pgfsetdash{{1pt}{1pt}}{0pt}% dashing cannot work correctly in all situation this way
    \pgfusepath{stroke}
   }
\pgfdeclarepatternformonly[\hatchspread,\hatchthickness,\hatchshift,\hatchcolor]% variables
   {nel}% name
   {\pgfqpoint{\dimexpr-2\hatchthickness}{\dimexpr-2\hatchthickness}}% lower left corner
   {\pgfqpoint{\dimexpr\hatchspread+2\hatchthickness}{\dimexpr\hatchspread+2\hatchthickness}}% upper right corner
   {\pgfqpoint{\dimexpr\hatchspread}{\dimexpr\hatchspread}}% tile size
   {% shape description
    \pgfsetlinewidth{\hatchthickness}
    \pgfpathmoveto{\pgfqpoint{\dimexpr\hatchshift-0.15pt}{-0.15pt}}
    \pgfpathlineto{\pgfqpoint{\dimexpr\hatchspread+0.15pt}{\dimexpr\hatchspread-\hatchshift+0.15pt}}
    \ifdim \hatchshift > 0pt
      \pgfpathmoveto{\pgfqpoint{-0.15pt}{\dimexpr\hatchspread-\hatchshift-0.15pt}}
      \pgfpathlineto{\pgfqpoint{\dimexpr\hatchshift+0.15pt}{\dimexpr\hatchspread+0.15pt}}
    \fi
    \pgfsetstrokecolor{\hatchcolor}
    \pgfusepath{stroke}
   }
\pgfdeclarepatternformonly[\hatchspread,\hatchthickness,\hatchshift,\hatchcolor]% variables
   {neld}% name
   {\pgfqpoint{\dimexpr-2\hatchthickness}{\dimexpr-2\hatchthickness}}% lower left corner
   {\pgfqpoint{\dimexpr\hatchspread+2\hatchthickness}{\dimexpr\hatchspread+2\hatchthickness}}% upper right corner
   {\pgfqpoint{\dimexpr\hatchspread}{\dimexpr\hatchspread}}% tile size
   {% shape description
    \pgfsetlinewidth{\hatchthickness}
    \pgfpathmoveto{\pgfqpoint{\dimexpr\hatchshift-0.15pt}{-0.15pt}}
    \pgfpathlineto{\pgfqpoint{\dimexpr\hatchspread+0.15pt}{\dimexpr\hatchspread-\hatchshift+0.15pt}}
    \ifdim \hatchshift > 0pt
      \pgfpathmoveto{\pgfqpoint{-0.15pt}{\dimexpr\hatchspread-\hatchshift-0.15pt}}
      \pgfpathlineto{\pgfqpoint{\dimexpr\hatchshift+0.15pt}{\dimexpr\hatchspread+0.15pt}}
    \fi
    \pgfsetstrokecolor{\hatchcolor}
    \pgfsetdash{{1pt}{1pt}}{0pt}% dashing cannot work correctly in all situation this way
    \pgfusepath{stroke}
   }
\begin{acronym}
	\acro{MPC}{Model predictive control}
	\acro{DMPC}{Distributed model predictive control}
	\acro{LSS}{large-scale system}
	\acro{mRPI}{minimal Robust Positively Invariant}
	\acro{QP}{Quadratic Programming}
	\acro{RCI}{Robust Control Invariant}
	\acro{RPI}{Robust Positive Invariant}
	\acro{PnP}{Robust Positive Invariant}
	\acro{OCP}{Optimal Control Problem}		
\end{acronym}

\newif\ifproves
\provestrue

\begin{document}
    \begin{frontmatter}
		\title{Coalitional predictive control: consensus-based coalition forming with robust regulation}
		\author[Paestum]{Pablo R. Baldivieso-Monasterios}\ead{p.baldivieso@sheffield.ac.uk}
		\author[Paestum]{Paul A. Trodden}\ead{p.trodden@sheffield.ac.uk}
		\address[Paestum]{Department of Automatic Control \& Systems Engineering, University of Sheffield, Mappin Street, Sheffield S1 3JD, UK}
        \begin{abstract}
          This paper is concerned with the problem of controlling a
          system of constrained dynamic subsystems in a way that balances the
          performance degradation of decentralized control with the
          practical cost of centralized control. We propose a
          coalitional control scheme in which controllers of
          subsystems may, as the need arises, group
          together into coalitions and operate as a single entity. The
          scheme employs a robust form of distributed model predictive
          control for which recursive feasibility and stability are
          guaranteed, yet---uniquely---the reliance on robust
          invariant sets is merely implicit, thus enabling
          applicability to higher-order systems. The robust control
          algorithm is combined with an algorithm for coalition
          forming based on consensus theory and potential games; we
          establish conditions under which controllers reach a
          consensus on the sets of coalitions. The recursive
          feasibility and closed-loop stability of the overall
          time-varying coalitional control scheme are established
          under a sufficient dwell time, the existence of which is
          guaranteed.
        \end{abstract}
    \end{frontmatter}
\section{Introduction}
\label{sec:introduction}
Distributed and decentralized forms of model predictive control (MPC)
have attracted significant attention as techniques for controlling
large-scale constrained systems. Many proposals have been made,
differing according to the nature or source of the coupling between
subsystems, and the algorithmic approach taken to coordinate control
actions~\citep{Scattolini09,MN14,CSM+13}.

A near-ubiquitous assumption in distributed MPC is that the
system is initially \emph{partitioned} into subsystems coupled
via dynamics, constraints or objectives. A typical approach then
assigns an MPC controller to each subsystem, and focuses on what
communication is needed, or assumptions necessary, to ensure
system-wide constraint satisfaction, stability and
optimality. Depending on the partition, however, the degree and
strength of coupling may change; indeed, it is well known that the
system partition has fundamental implications for many aspects
of control system design and operation including, \emph{inter alia},
controllability and observability, dimensionality and complexity,
communication, stability and
performance~\citep{Siljakbook}.

A question that naturally arises, therefore, is what is the best
choice of system partition for a system that is to be controlled by
distributed or decentralized MPC? There is relatively little on this
in the literature: early work by \citet{MS03} proposed to select an
optimal partition online by minimizing an unconstrained open-loop
performance index; more recent
contributions~\citep{Pourkargar2017a,Zheng2018}, considering constraints,
have used community detection algorithms to decompose, offline, the
system-wide optimal control problem into problems with minimal overlap
and weak interactions. Similarly, \citep{BG2017,BG2019} focus on finding offline a suitable partition based on different criteria, for example minimization of communication requirements, and topologies that  facilitate the computation of game theoretic metrics.

A contemporaneous and relevant development is the emergence of
\emph{coalitional control} schemes~\citep{FMC17}. Such schemes aim to
design control strategies that optimize (on-line) the trade-off
between control performance, complexity, and communication. This is
achieved by controllers acting cooperatively or independently at
different times. This is the idea pursued in the current paper.

The coalitional control literature first considered an unconstrained
linear quadratic (LQ) setting and focused on analysing the benefits of
controlling subsystems in coalitions. Tools from cooperative game
theory---the Shapley value~\citep{Muros2018}, Harsanyi power
solutions~\citep{MAM+17}, and the Banzhaf value~\citep{MAM+17a}---have
been applied and studied. When constraints are present, however,
unconstrained LQ control may be deficient;~\citet{FDM+18} therefore
proposed a coalitional MPC scheme wherein constraints are handled
naturally, and the system partition is determined from bargaining
between predictive controllers. Under assumptions of recursive
feasibility and weak coupling, input-to-state (ISS) stability was
established regardless of when and which coalitions were formed.

The problem of guaranteeing recursive feasibility and closed-loop
stability in coalitional MPC is actually non-trivial. The main issue
is indeed that the time-varying coalitional system is a switched or
switching system for which feasibility and stability are not naturally
maintained. Additionally, coalitional MPC inherits and exacerbates the
fundamental challenges of distributed MPC for dynamically coupled
subsystems: in order to \emph{guarantee} feasibility and stability,
the control algorithm must either rely on iteration between
controllers at each sampling time~(\eg~\citep{VHR+08}), or use
techniques from robust MPC (\eg~\citep{FS12,TM17}), even though the
underlying control problem is a nominal one. The latter family of
approaches are iteration-free but bring their own challenges in the
coalitional setting: firstly, the dynamics of a coalition may be of
high order, even if the constituent subsystem dynamics are of low
order; secondly, the coalitions will vary, both in size and in
membership, over time. Both of these features render impractical a
control approach based on robust invariant sets, since these are
prohibitively difficult to compute---even offline---for systems of
anything other than low order.

With these challenges in mind, the contributions of this paper are
three-fold:
\begin{itemize}
\item In Section~\ref{sec:dmpc_fixed_partition}, we present a
  distributed scheme for time-invariant coalitions of subsystems with
  guarantees of robust recursive feasibility and stability, despite the a-priori unknown disturbances arising from interactions, yet minimal
  reliance on invariant sets. The proposed scheme, first
  developed in~\citep{BHT17}, employs two MPC controllers for each
  subsystem. Crucially, the design and formulation of the MPC
  problems, and the invariance-inducing control law, does not require
  the explicit characterization of a robust invariant set, but relies
  only implicitly on the existence of one.
\item We propose, in Section~\ref{sec:switching_between_partitions}, a
  scheme for selecting partitions online using consensus
  optimization. Subsystems optimize their opinion and reach a
  consensus on what the system partition should be at the current
  state. The choice of consensus objective function is shown to be a
  potential function, and the consensus algorithm inherits strong
  properties from potential games, including finite-iteration
  convergence.
\item In Section~\ref{sec:docp_implementation}, we combine the robust
  control and consensus-based partition selection algorithms to
  produce a time-varying coalitional control scheme wherein controller
  re-design, in response to the system having re-organized into new
  coalitions, requires the solving of a linear programming problem.
  We study the properties of the time-varying system, and establish
  conditions under which it is feasible and stable.
\end{itemize}

Section~\ref{sec:preliminaries} defines the problem. The
results are illustrated in Section~\ref{sec:simulations}. Proofs are given in the Appendix.

\emph{Notation and basic definitions}: $\mbb{I}_{\geq 0}$ and
$\mbb{I}_{>0}$ are the sets of non-negative and positive integers;
$\mbb{I}_{a:b}$ is the set of integers between $a < b$.
$\mbb{R}_{\geq 0}$ and $\mbb{R}_{>0}$ are the sets of non-negative and
positive real numbers. $|x|$ denotes the $\ell_2$-norm, $\| x \|_2$,
of a vector $x \in \mbb{R}^n$. A C-set is a compact and convex set
containing the origin, while a PC-set is a C-set with the origin in
its interior. For two sets $A$ and $B$, the \emph{Minkowski sum} is
$A\oplus B =\{a + b : a\in A, b\in B\}$, and the \emph{Pontryagin
  difference} is
$A\ominus B =\{a : a + b\in A, \ \forall b\in B\}$. A set
$\mc{R}$ is \emph{robust control invariant} (RCI) for a system
$x^+=f(x,u,w)$ and constraints $(\mbb{X},\mbb{U},\mbb{W})$
if (i) $\mc{R}\subset\mbb{X}$ and (ii)$\forall x\in\mc{R}$, $\exists u = \mu(x) \in \mbb{U}$ such that $x^+=f(x,u,w)\in\mc{R}$,
$\forall w\in\mbb{W}$; the control law $u = \mu(x)$ is said to be
\emph{invariance inducing} over the set $\mc{R}$. The origin is
locally stable for a discrete-time system $x^+ = f(x)$ if, for any
$\epsilon > 0$, there exists a $\delta > 0$ such that
$\left| x(0) \right| \leq \delta$ implies $\forall k \in \mbb{I}_{>0}$, 
$\left| x(k) \right| \leq \epsilon $; if,
in addition, for any $x(0) \in X$, $x(k) \to 0$ as $k \to \infty$ then
the origin is asymptotically stable with region of attraction $X$; if
$\left| x(k) \right| \leq c \gamma^k \left| x(0) \right|$ for all
$x(0) \in X$, where $c >0 $ and $\gamma \in (0,1)$, then the origin is
exponentially stable with region of attraction $X$. A function
$\alpha \colon \mbb{R}_{\geq 0} \to \mbb{R}_{\geq 0}$ is said to
be a $\mc{K}$-function if it is continuous, strictly increasing, and
has $\alpha(0) = 0$.

\section{Problem statement and preliminaries}
\label{sec:preliminaries}

\subsection{The system and its partition into subsystems}
\label{sec:lss}
We consider the problem of controlling a discrete-time, linear time-invariant system
\begin{equation}
		{x}^+=A{x}+B{u},
\label{eq:large_scale}
\end{equation}
where ${x}\in\Rset^n$, ${u}\in\Rset^m$ are the state and control
input, and ${x}^+$ is the state at the next instant of time. We
consider that a basic \emph{partitioning}
of~\eqref{eq:large_scale} into a number, $M$, of independently
actuated subsystems is known. The dynamics of subsystem
$i \in \mc{M} \triangleq \{1,\ldots,M\}$ are
\begin{equation*}
  x_i^+=A_{ii}x_i+B_iu_i+ w_i \ \text{where} \ w_i \triangleq \sum\limits_{\inN{j}{i}} A_{ij}x_j,
%\label{eq:subsystem_partition}
\end{equation*}
and $x_i\in\Rset^{n_i}$, $u_i\in\Rset^{m_i}$ are the state and input of $i \in \mc{M}$, with $x = (x_1,\dots,x_M)$,
$u = (u_1,\dots,u_M)$ the aggregate state and input respectively.  The set of \emph{neighbours} of subsystem $i$ is $\mc{M}_i \triangleq \bigl\{ j\in\mc{M} \setminus \{i\} :  A_{ij} \neq 0\bigr\}.$
%
%\begin{equation*}
%
%  \end{equation*}
  %
  \begin{assumption}[Controllability]
For each $i\in\mc{M}$ the pair $(A_{ii},B_i)$ is controllable. 
	\label{assump:controllability}
\end{assumption}

The system is constrained via local, independent constraints on the
states and inputs of each subsystem,
\ie~$x_i \in \mbb{X}_i$, $u_i \in \mbb{U}_i$ for subsystem $i$.
\begin{assumption}[Constraint sets]
  The sets $\mbb{X}_i\subset\Rset^{n_i}$ and
  $\mbb{U}_i\subset\Rset^{m_i}$ are PC-sets.
  \label{assump:constraints}
\end{assumption}

\subsection{Coalitions of subsystems and partitions of the system}
\label{sec:coals}

The setting of the paper is to consider that subsystems may grouped
together into, and controlled as, \emph{coalitions}.
\begin{definition}[Coalition of subsystems]
A \emph{coalition} of subsystems is a non-empty subset of $\mc{M}$.
\end{definition}
The idea is that each coalition of subsystems operates and is
controlled as a single entity; a coalitional controller replaces (or
coordinates) the local subsystem controllers. Viewed differently, the
grouping of the subsystems into coalitions induces an alternative
partitioning of the system.
\begin{definition}[Partition of the system]
  A \emph{partition} of the system is an arrangement of the $M$ subsystems into $C \leq M$ coalitions: formally, the partition of $\mc{M} = \{1,\ldots,M\}$ is the set $\mc{C} = \{1,\dots,C\}$, satisfying the following properties:
  \begin{enumerate}
  \item Coalition $c \in \mc{C}$ contains subsystems $c \subseteq \mc{M}$; the cardinality of $c$ is $M_c$.
  \item Coalitions are non-overlapping: $c \cap d = \emptyset$ for all $c \neq d$ and $c,d\in\mc{C}$.
    \item Coalitions cover the set of subsystems: $\bigcup_{c\in \mc{C}} c = \mc{M}$.
    \end{enumerate}
\label{def:partition}
\end{definition}

These definitions include the trivial cases of (i) a single, grand coalition of all subsystems ($C = 1$, $c_1 = \mc{M}$) (the \emph{centralized partition}) and (ii) the basic partitioning of the system, in which each subsystem is a coalition ($C = M$, $\mc{C} = \mc{M}$, $c = \{i\}$ for each $i \in \mc{M}$) (the \emph{decentralized partition}). The set of all possible partitions is
\begin{equation*}
\Pi_{\mc{M}} \triangleq \{\mc{C} \st \mc{C} \ts{ is a partition of } \mc{M}\}.
\end{equation*}
Given a partition $\mc{C}$, the state and input of coalition $c$ are,
respectively, $x_c = (x_i)_{i \in c}$ and
$u_c = (u_i)_{i \in c}$~\footnote{Our intention is to make the
  notation as simple as possible by employing a single subscript to
  denote both a variable of a subsystem and a variable of a
  coalition.}. The dynamics of coalition $c$ are
\begin{equation*}
  x_c^+=A_{cc}x_{c}+B_cu_{c} + w_c,
  %\label{eq:coal_dyn}
\end{equation*}
where the matrices $A_{cc}$ and $B_c$ contain, as sub-blocks, the matrices of subsystems within the coalition:
$A_{cc} = [A_{ij}]_{i,j\in c}$,
$B_c=\diag{B_{i}}_{i\in c}$. Similar to the system basic partition, the coalitions remain coupled via
their dynamics: coalition $c$ is coupled with coalition $d$ via the
matrices $A_{cd} = \bigl[A_{ij}\bigr]_{i\in {c},j\in{d}, d \neq c}$
so that
\begin{equation*}
w_c \triangleq \sum_{d \in \mc{M}_c} A_{cd}x_{d} \ \text{where} \ \mc{M}_c \triangleq \bigl\{ d\in \mc{C}\setminus\{c\} : A_{cd} \neq 0\bigr\}.
\end{equation*}

\begin{assumption}
For any partition $\mc{C} \in \Pi_\mc{M}$, each pair $(A_{cc},B_c)$, for $c \in \mc{C}$, is controllable.
  \end{assumption}

\subsection{Coalitional control problem}
\label{controlprob}
The aim is to solve the following optimal control problem: from a
state $x(0)$, determine the control policy and coalitional policy that
minimizes the bi-criteria cost
\begin{equation}
	 \underbrace{\sum_{k=0}^\infty  x^\top(k) Q x(k) + u^\top(k) R u(k) }_{V^{\infty}(x(0),\mb{u}(0))} + J(\mc{C}(k),x(k))
	\label{eq:infine_cost}
\end{equation}
with $Q \triangleq \diag{Q_1,\dots,Q_M}$,
$R \triangleq \diag{R_1,\dots,R_M}$, while satisfying constraints
$x(k) \in \mbb{X} \triangleq \mbb{X}_1 \times \dots \times \mbb{X}_M$,
$u(k) \in \mbb{U} \triangleq \mbb{U}_1 \times \dots \times \mbb{U}_M$
for $k \in \mbb{I}_{\geq 0}$. The term $J(\mc{C},x)$ is
  supposed to measure the \emph{practical} or \emph{operating} cost of
  controlling subsystems in coalitions: it may include, for instance,
  costs on communication, computation and complexity.
\begin{assumption}[Positive definite stage cost]\label{assump:pd}
$Q_i$ and $R_i$ are, for each $i \in \mc{M}$, positive definite matrices.
\end{assumption}
The idea is to determine the infinite-horizon control sequence
$\mb{u}(0) = \bigl\{u(1),u(2),\dots\bigr\}$ and partition sequence
$\left\{\mc{C}(0), \mc{C}(1), \dots \right\}$ that minimizes this
system-wide joint cost on regulation performance and practical
operation. (In contrast, the coalitional MPC scheme of~\citet{FDM+18}
aims to minimize individual subsystem performance costs by using
coalitions and game-theoretical measures to allocate payoffs.) It has
been shown, via a range of applications, that there is a potential
benefit to performance of employing different coalitions over
time~\citep{MMJ+14,MMA+17,MMA+14}. However, the optimal control
problem is generally intractable---even when $J(\mc{C})$ is well
defined---as it is an infinite-dimensional combinatorial optimization
problem. Thus, in the sequel we propose a suboptimal way to solve this
problem while achieving guarantees of constraint satisfaction and
stability.

\section{Robust MPC for time-invariant coalitions}
\label{sec:dmpc_fixed_partition}

We first consider the scenario where the set of subsystems $\mc{M}$
are arranged into a collection of fixed coalitions
$\{c_1,c_2,\dots,c_C\}$. The aim is for each coalition, acting as a
single entity, to regulate its combined state to the origin, while
respecting constraints. To this end, each coalition is equipped with a
model predictive controller.\footnote{The MPC problem for a coalition
  can be solved by a single agent in the coalition (a leader), or
  distributed among several members, but these details are beyond the
  scope of this paper.} Owing to the presence of dynamic coupling
between coalitions, manifested as the disturbance
$w_c = \sum_{d \in \mc{M}_c} A_{cd} x_d$ for coalition $c$,
consideration needs to be given to handling interactions adequately in
order to achieve constraint satisfaction and
stability~\citep{Scattolini09}.

Among the numerous DMPC schemes,
algorithms based on robust techniques~\citep{MSR05} have the advantage
of achieving feasibility and stability guarantees without relying on
inter-agent iterations and negotiation~\citep{FS12,RF12,TM17}. The
fundamental ingredient for such schemes is the availability of an RCI
set, $\mc{R}_c$, for the uncertain dynamics of each coalition that
arise when the state interaction is treated as a disturbance (\ie~
$x_c^+ = A_{cc} x_c + B_{c}u_c + w_c$ with
$w_c = \sum_{d \in \mc{M}_c} A_{cd} x_d$ a-priori unknown) along with
its invariance-inducing control law $\tilde{\kappa}_c(\cdot)$. In the
simplest implementation, the set $\mc{R}_c$ is used to tighten the
constraint sets, \ie~as $\mbb{X}_c \ominus \mc{R}_c$ and
$\mbb{U}_c \ominus \tilde{\kappa}_c(\mc{R}_c)$, in a \emph{nominal} MPC
problem that employs the disturbance-free prediction model
$\bar{x}_c^+ = A_{cc} \bar{x}_c + B_c \bar{u}_c$ involving
\emph{nominal} prediction variables $\bar{x}_c$ and $\bar{u}_c$. The
composite control law
$u_c = \bar{\kappa}_c(\bar{x}_c) + \tilde{\kappa}_c(x_c -
\bar{x}_c)$---where the first term is the implicit control law arising
from the nominal MPC---bounds the mismatch between true variables
$(x_c, u_c)$ and nominal variables $(\bar{x}_c, \bar{u}_c)$, and
ensures recursive feasibility of the MPC problems and stability of the
closed-loop system.

As explained in the Introduction, however, the features of the
coalitional control problem render these approaches impractical. To
address this challenge, we adopt therefore the ``nested'' robust approach
initially developed in~\citep{BHT17}. This approach replaces the
ancillary robust control law---which usually requires knowledge of
$\mc{R}_c$---with a secondary MPC controller. Constraint restrictions
in the primary MPC formulation are achieved via simple scalings of
$\mbb{X}_c$ and $\mbb{U}_c$ rather than the exact restrictions. We find
that the closed-loop properties of the scheme rely on the
\emph{implicit} existence of an RCI set, with the implication for
design and implementation that there is no need to either explicitly
characterize or compute the RCI set, or impose it anywhere in the MPC
constraints.  Consequently, the dependency on invariant sets is
minimized, while stability and feasibility guarantees are retained,
making the approach more suitable for higher-order dynamics.

  \subsection{Primary MPC controller for coalition $c \in \mc{C}$}

  The primary controller, following conventional tube-based MPC,
  employs a nominal prediction model along with simple
  constraint restrictions. For coalition
  $c \in \mc{C}$ with (nominal) state $\bar{x}_c$, the optimal control problem is
\begin{equation*}
  \bar{\mbb{P}}_{c}(\bar{x}_{c}) \colon  \min_{\bar{\mb{u}}_{c}} \bigl\{ V^N_{c}(\bar{x}_c,\bar{\mb{u}}_c) : \bar{\mb{u}}_c \in \bar{\mc{U}}^N_{c}(\bar{x}_{c}) \bigr\}
\end{equation*}
% \bar{V}^0_c(\bar{x}_c) =
where the decision variable
$\bar{\mb{u}}_c \triangleq \bigl\{ \bar{u}_c(0), \dots, \bar{u}_c(N-1)
\bigr\}$, $V^N_c$ is the finite-horizon regulation
cost\footnote{$\bar{x}_c(j)$ denotes the prediction of state
  $\bar{x}_c$ at prediction step $j$, starting from $\bar{x}_c(0) = \bar{x}_c$, the current measurement of $\bar{x}_c$.}
\begin{equation*}
	V^N_{c}(\bar{x}_c,\bar{\mb{u}}_{c}) = \sum_{j=0}^{N-1} \bar{x}_c^\top(j) Q_c \bar{x}_c(j) + \bar{u}_c^\top (j) R_c \bar{u}_c(j),
\end{equation*}
with $Q_c \triangleq \diag{Q_i}_{i \in c}$, $R_c \triangleq \diag{R_i}_{i \in c}$
and $\bar{\mc{U}}^N_{c}(\bar{x}_{c})$ is defined by the following constraints for $j\in \mbb{I}_{0:N-1}$:
%
%\begin{subequations}
\begin{align*}
	\bar{x}_{c}(0) &= \bar{x}_c, \\ %\label{eq:coal_cons_init}\\
	\bar{x}_{c}(j+1) &= A_{cc} \bar{x}_{c}(j) + B_c\bar{u}_{c}(j), \\%\label{eq:coal_cons_dyn}\\
	\bar{x}_{c}(j) &\in \alpha_c^x \mbb{X}_{c},\\%\label{eq:coal_cons_x}\\
	\bar{u}_{c}(j) &\in \alpha_c^u \mbb{U}_{c},\\%\label{eq:coal_cons_u}\\
	\bar{x}_{c}(N) &= 0,%\label{eq:coal_cons_xf}
\end{align*}
%\end{subequations}
%
where $\mbb{X}_c \triangleq \prod_{i \in c} \mbb{X}_i$ and
$\mbb{U}_c \triangleq \prod_{i \in c} \mbb{U}_i$. The simple
choice of the origin as terminal set is to facilitate the
applicability to higher-order dynamics. Selection of the scaling
parameters $\alpha_c^x, \alpha_c^u \in (0,1)$ is described later.

Problem $\bar{\mbb{P}}_c(\bar{x}_c)$ is a finite-horizon
approximation to the coalition $c$'s share of the infinite-horizon
problem~\eqref{eq:infine_cost}, omitting the cost $J(\mc{C})$. Solving
this problem yields the sequence of nominal control actions $\bar{\mb{u}}_c^0(\bar{x}_c)\triangleq\{\bar{u}^0_c(0;\bar{x}_c),\ldots,\bar{u}_c^0(N-1;\bar{x}_c)\}$. Taking
the first term of the sequence and applying it to coalition $c$ defines the implicit feedback law $\bar{\kappa}_c(\bar{x}_c) = \bar{{u}}_c^0(0;\bar{x}_c)$. The primary problems are solved in parallel by each coalition, and each then communicates the associated optimized state sequence $\bar{\mb{x}}^0_c(\bar{x}_c)$, to its neighbours $d \in \mc{M}_c$.

\subsection{Secondary MPC controller for coalition $c \in \mc{C}$}

Having received $\bar{\mb{x}}^0_d(\bar{x}_d)$ from neighbouring coalitions
$d \in \mc{M}_c$, the controller for coalition $c$ solves a secondary
MPC problem employing a refined prediction model 
\begin{equation}\label{eq:secnomsys}
\hat{x}^+_c = A_{cc}\hat{x}_{c}+B_c\hat{u}_{c} + \bar{w}_{c},
\end{equation}
where
$\bar{w}_{c} = \sum_{d\in\mc{M}_c} A_{cd}\bar{x}_d$
is the \emph{planned} disturbance using the primary information
obtained from the neighbours; this information forms the $(N+1)$-length sequence of future disturbances, $\bar{\mb{w}}_c \triangleq \{ \bar{w}_c(0), \bar{w}_c(1), \dots, \bar{w}_c(N) \},$
%
%
%\begin{equation*}
%
%  \end{equation*}
%
where $\bar{w}_c(j) = \sum_{d\in\mc{M}_c} A_{cd}\bar{x}_d^0(j;\bar{x}_d)$.

The aim of the secondary controller is to design perturbations to the nominal
$\bar{\mb{u}}^0_c(\bar{x}_c)$ in order to handle the planned
interactions. We define the error variables
$\bar{e}_c \triangleq \hat{x}_c - \bar{x}_c$ and
$\bar{f}_c \triangleq \hat{u}_c - \bar{u}_c$.  The secondary problem is then
\begin{equation*}
\hat{\mbb{P}}_c(\bar{e}_c;\bar{\mb{w}}_c) \colon \min_{\bar{\mb{f}}_c} \bigl\{ V^H_{c}(\bar{e}_c,\bar{\mb{f}}_c) : \bar{\mb{f}}_c \in \bar{\mc{F}}^H_{c}(\bar{e}_{c};\bar{\mb{w}}_c)\bigr\}
%\label{eq:coal_anc_ocp} \hat{V}^0_c(\bar{e}_c; \bar{\mb{w}}_c) = 
\end{equation*}
where the cost function has, for simplicity but not necessity, the
same structure as the one in the primary problem (albeit in terms of
variables $\bar{e}_c$ and $\bar{\mb{f}}_c$ and a horizon $H$), and the
set $\bar{\mc{F}}^H_{c}(\bar{e}_{c};\bar{\mb{w}}_c)$ is defined by the
following constraints for $j \in \mbb{I}_{0:H-1}$:
%
%\begin{subequations}
\begin{align*}
	\bar{e}_{c}(0) &= \bar{e}_c,\\% \label{eq:coal_cons_anc_init}\\
	\bar{e}_{c}(j+1) &= A_{cc} \bar{e}_{c}(j) + B_c\bar{f}_{c}(j) + \bar{w}_c(j),\\%\label{eq:coal_cons_anc_dyn}\\
	\bar{e}_{c}(j) &\in \beta_c^x \mbb{X}_{c},\\%\label{eq:coal_cons_anc_x}\\
	\bar{f}_{c}(j) &\in \beta_c^u \mbb{U}_{c},\\%\label{eq:coal_cons_anc_u}\\
	\bar{e}_{c}(H) &= 0.%\label{eq:coal_cons_anc_xf}
\end{align*}
%\end{subequations}
%
Similar to the primary problem, the constraint sets are scaled
versions of the original sets, albeit with different scaling
factors. The horizon of this problem is $H$; since $\bar{w}_c(N)=0$,
then setting $H \geq N + 1$ will ensure that the disturbance is dealt
with during the first $N$ steps of the predictions, with the remaining
$H - N$ steps allowing the predicted error to be driven to zero, as
required by the terminal constraint. The solution of this problem is the sequence of controls
\begin{equation*}
  \bar{\mb{f}}_c^0(\bar{e}_c;\bar{\mb{w}_c})\triangleq \{\bar{f}^0_c(0;\bar{e}_c,\bar{\mb{w}}_c),\ldots,\bar{f}_c^0(H-1; \bar{e}_c, \bar{\mb{w}}_c)\}.
\end{equation*}
Since $e_c = \hat{x}_c - \bar{x}_c$ and $f_c = \hat{u}_c - \bar{u}_c$,
selecting the first element of
$\bar{\mb{f}}_c^0(\bar{e}_c;\bar{\mb{w}}_c)$ and adding to
$\bar{u}_c^0(\bar{x}_c)$ yields the two-term control law
\begin{equation*}
  \bar{\kappa}_c (\bar{x}_c) + \hat{\kappa}_c(\bar{e}_c;\bar{\mb{w}}_c) = \bar{u}_c^0(0;\bar{x}_c) + \bar{f}_c^0(0;\bar{e}_c,\bar{\mb{w}}_c)
  \end{equation*}
  that, under suitable conditions, stabilizes
  $\hat{x}_c = A_{cc}\hat{x}_c + B_c\hat{u}_c + \bar{w}_c$.

\subsection{Overall robust controller and algorithm}
  
Closing the loop with
$u_c =\bar{\kappa}_c (\bar{x}_c) +
\hat{\kappa}_c(\bar{e}_c;\bar{\mb{w}}_c) $ does not, however,
guarantee constraint satisfaction, feasibility and
stability for the true coalition dynamics
$x_c^+ = A_{cc} x_c + B_cu_c + w_c$, because part of the interaction
is still neglected: the true disturbance is
$w_c = \sum_{d \in \mc{M}_c} A_{cd} x_d$ and not the planned
one, $\bar{w}_c = \sum_{d \in \mc{M}_c} A_{cd}\bar{x}_d$, used
for predictions in the secondary MPC.

The control law is, therefore, completed with a final term
$\tilde{\kappa}_c(\hat{e}_c)$---the requirements on which are given in
the next section---that acts on the \emph{unplanned} error
$\hat{e}_c \triangleq x_c - \hat{x}_c$ that arises from the unplanned,
residual disturbance $\hat{w}_c \triangleq w_c - \bar{w}_c$. The total error is $e_c = \bar{e}_c + \hat{e}_c = x_c - \bar{x}_c$. Since $x_{c} = \bar{x}_{c} + \bar{e}_{c} + \hat{e}_{c}$ and
$u_{c} = \bar{u}_{c} + \bar{f}_{c} + \hat{f}_{c}$, the resulting
three-term policy defines a feedback control law on the true state $x_c$:
\begin{equation}
u_c =\kappa_{c}(x_{c}) \triangleq \bar{\kappa}_{c}(\bar{x}_{c}) + \hat{\kappa}_{c}(\bar{e}_{c};\bar{\mb{w}}_{c}) + \tilde{\kappa}_{c}(\hat{e}_{c}).
	\label{eq:control_law_nested}
\end{equation}
This three-term control law is employed in Algorithm~\ref{alg:nedmpc}.

\begin{alg}[MPC for coalition $c$]~
  
\textbf{Initial data}: Sets $\mbb{X}_{c}$, $\mbb{U}_{c}$, $\mc{M}_{c}$; matrices $A_{cd}$ for $d \in \mc{M}_c$; constants $\alpha_{c}^x$, $\alpha_{c}^u$, $\beta_{c}^x$,$\beta_{c}^u$; states $\bar{x}_{c}(0) = x_{c}(0)$, $\bar{e}_{c} = 0$, $\bar{\mb{w}}_{c} = \mb{0}$, $\hat{V}_{c} = +\infty$.

\textbf{Online Routine}:
\begin{enumerate}
\item\label{step:solve_main} At time $k$, controller state
  $\bar{x}_c$, solve $\bar{\mbb{P}}_{c}(\bar{x}_{c})$ to obtain
  $\bar{\mb{u}}_{c}^0$ and $\bar{\mb{x}}_{c}^0$.
\item\label{step:transmit} Transmit $\bar{\mb{x}}_{c}^0$ to 
  $d \in \mc{M}_{c}$; having received $\bar{\mb{x}}_{d}^0$ from $d \in \mc{M}_c$, compute $\bar{\mb{w}}^0_c = \sum_{d\in\mc{M}_c} A_{cd} \bar{\mb{x}}_{d}^0$.
\item\label{step:solve_anc} At controller state $\bar{e}_{c}$, attempt to solve $\hat{\mbb{P}}_{c}(\bar{e}_{c};\bar{\mb{w}}^0_{{c}})$ to obtain $\bar{\mb{f}}_{c}^0$: if the problem is feasible and $\hat{V}^0_{c}(\bar{e}_{c}, \bar{\mb{f}}^0_{{c}}) \leq \hat{V}_{c}$, then set $\bar{\mb{w}}_{c} = \bar{\mb{w}}^0_{c}$ and $\hat{V}_{c} = \hat{V}_{c}^0(\bar{e}_{c},\bar{\mb{f}}^0_{c})$; otherwise, solve $\hat{\mbb{P}}_{c}(\bar{e}_{c};\bar{\mb{w}}_{{c}})$ for $\bar{\mb{f}}_{c}^0$.
\item\label{step:law} Measure plant state $x_{c}$, calculate $\hat{e}_{c} = x_{c} - \bar{x}_{c} - \bar{e}_{c}$, and apply $u_{c} = \bar{u}_{c}^0 + \bar{f}_{c}^0 + \tilde{\kappa}_c(\hat{e}_{c})$.
\item\label{step:update} Update controller states as $\bar{x}_{c}^+ = A_{cc}\bar{x}_{c}+B_c\bar{u}^0_{c}$ and $\bar{e}_{c}^+ = A_{cc}\bar{e}_{c}+B_c\bar{f}_{c}^0+\bar{w}_{c}$ (where $\bar{w}_c$ is the first element in $\bar{\mb{w}}_c$), $\bar{\mb{w}}^+_{c} = \{\bar{w}_{c}(1),\dots,\bar{w}_{c}(N),0\}$, and $\hat{V}^{+}_{c} = \hat{V}_{c} - [\bar{e}_{c}^\top Q_c \bar{e}_c + \bar{f}^{0\top}_{c}R_c \bar{f}^0_c]$.

\item Wait one time step; set $k=k+1$, $\bar{x}_{c} = \bar{x}_{c}^+$, $\bar{e}_{c} = \bar{e}_{c}^+$, $\bar{\mb{w}}_{c} = \bar{\mb{w}}_{c}^+$, $\hat{V}_{c} = \hat{V}_{c}^{+}$, and go to Step~\ref{step:solve_main}.

\end{enumerate}
\label{alg:nedmpc}
\end{alg}

\subsection{Closed-loop properties}

Recursive feasibility and stability of the algorithm were
established in~\citet{BHT17}, and are here
tailored to the coalitional setting. The procedure for the
design of the final term in the control law and the constraint scaling parameters, given in~\ref{sec:design}, is assumed to terminate
having met the following assumptions. 

\begin{assumption}
  \label{assump:invariance_R_h}
  The control law $\hat{f}_c = \tilde{\kappa}_c(\hat{e}_c)$ is
  invariance inducing over a set ${\mc{R}}_c$ that is RCI for the
  system $\hat{e}_c^+ = A_{cc}\hat{e}_c + B_c\hat{f}_c + \hat{w}_c$
  and constraint set
  $(\xi_c^x\mbb{X}_c, \xi_c^u\mbb{U}_c, \hat{\mbb{W}}_c)$, for some
  $\xi_c^x\in [0,1)$ and $\xi_c^u\in [0,1)$, and where
  $\hat{\mbb{W}}_c \triangleq \bigoplus_{d \in \mc{M}_c}
  (1-\alpha^x_d)A_{cd}\mbb{X}_d$.
\end{assumption}

 \begin{assumption}
  \label{assump:constraint_sat}
  The constants $(\alpha_c^x,\beta_c^x,\xi_c^x)$ and
  $(\alpha_c^u,\beta_c^u,\xi_c^u)$ satisfy
  $\alpha_c^x+\beta_c^x+\xi_c^x\leq 1$ and
  $\alpha_c^u+\beta_c^u+\xi_c^u\leq 1$.
  \end{assumption}

  To aid the
  statement of the results, we make the following definitions:
  $\bar{\mbb{W}}_c = \bigoplus_{d \in \mc{M}_c} \alpha_d^x A_{cd}
  \mbb{X}_d$ is the set of
  disturbances arising from admissible state predictions ($\bar{x}_d \in \alpha_d^x \mbb{X}_d$) for coalitions $d \in \mc{M}_c$; the set
  $\bar{\mc{W}}^N_c \triangleq \bar{\mbb{W}}_c \times \bar{\mbb{W}}_c
  \times \dots \times \bar{\mbb{W}}_c \times \{0\}$ corresponds to admissible state \emph{sequences}. Given a disturbance sequence
  $\bar{\mb{w}}_c = \{\bar{w}_c(0),\dots,\bar{w}_c(N-1),0\} \in
  \bar{\mc{W}}^N_c$,
  $\bar{\mb{w}}^+_c = \{\bar{w}_c(1),\dots,\bar{w}_c(N-1),0,0\}$ is
  the tail of that sequence plus a terminal zero. The domain of $\bar{\mbb{P}}_{c}(\bar{x}_{c})$ is $	\bar{\mc{X}}^N_{c} \triangleq \{ \bar{x}_{c} : \bar{\mc{U}}_{c}^N(\bar{x}_{c})\neq\emptyset\}$, while the corresponding domain of the secondary problem $\hat{\mbb{P}}_c(\bar{e}_c; \bar{\mb{w}}_c)$, which depends on the parameter $\bar{\mb{w}}_c$, is $	\bar{\mc{E}}^H_c(\bar{\mb{w}}_c) \triangleq \{ \bar{e}_c : \bar{\mc{F}}_c^H(\bar{e}_c;\bar{\mb{w}}_{c})\neq\emptyset\}$. Recursive feasibility for time-invariant coalitions is then established in the following proposition.
\begin{proposition}[Recursive feasibility]
    Suppose that Assumptions~\ref{assump:controllability}--\ref{assump:constraint_sat} hold. Then, for each coalition $c \in \mc{C}$:
\begin{enumerate}[(i)]
\item If $\bar{x}_{c} \in \bar{\mc{X}}_{c}^N$ then
  $A_{cc} \bar{x}_c + B_c \bar{\kappa}_c(\bar{x}_c) \in \bar{\mc{X}}_{c}^N$.
\item If $\bar{e}_{c} \in \bar{\mc{E}}_{c}^H(\bar{\mb{w}}_{c})$ for
  some $\bar{\mb{w}}_{c} \in \bar{\mc{W}}^N_{c}$, then
  $A_{cc} \bar{e}_c + B_c\hat{\kappa}_c(\bar{e}_c;\bar{\mb{w}}_c) + \bar{w}_c \in \bar{\mc{E}}_c^H(\bar{\mb{w}}^+_c)$, where $\bar{w}_c = \bar{\mb{w}}_c(0)$.
\item If $\bar{x}_{c}(0)=x_{c}(0)\in \bar{\mc{X}}^N_{c}$ then the
  coalition dynamics $x_{c}^+=A_{cc}x_{c}+B_cu_{c}+w_{c}$ under the control
  law
  $u_{c} = \bar{\kappa}_c(\bar{x}_c) +
  \hat{\kappa}_c(\bar{e}_c;\bar{\mb{w}}_c) +
  \tilde{\kappa}_c(\hat{e}_c)$ satisfy $x_c(k)\in\mbb{X}_{c}$ and
  $u_{c}(k)\in\mbb{U}_{c}$ for $k \in \mbb{I}_{\geq 0}$.
\end{enumerate}
\label{prop:feas}
\end{proposition}

Stability then follows under the following assumption, which ensures that once the state has entered the robust invariant set around the origin, the collection of invariance-inducing control laws then bring the state asymptotically to the origin.

\begin{assumption}[Decentralized stabilizability]\label{assump:decent}
  The control laws $u_c = \tilde{\kappa}_c(x_c)$, $c \in \mc{C}$, together asymptotically
  stabilize the system $x^+ = Ax+Bu$ in a neighbourhood $\mc{R}_\mc{C}$ of the origin.
\end{assumption}

\begin{theorem}[Stability]\label{thm:stab}
  Suppose that
  Assumptions~\ref{assump:controllability}--\ref{assump:decent}
  hold. Then, for each $c \in\mc{C}$, the origin is exponentially
  stable for the nominal coalition system
  $\bar{x}_c^+ = A_{cc} x_c + B_c \bar{\kappa}_c(\bar{x}_c)$ and
  asymptotically stable for the true coalition system
  $x_c^+ = A_{cc}x_c + B_c\kappa_c(x_c) + \sum_{d \in \mc{M}_c} A_{cd}
  x_d$. The region of attraction for $(\bar{x}_c, x_c)$ is
  $\bar{\mc{X}}^N_c \times \bar{\mc{X}}^N_c$.
\end{theorem}

Finally, we note some consequences of these results.

\begin{corollary}
  For each $c \in \mc{C}$, the sets $\bar{\mc{X}}_c^N$ and
  $\bar{\mc{X}}^{N-1}_c$ are positively invariant for the nominal
  dynamics
  $\bar{x}_c^+ = A_{cc} \bar{x}_c + B_c \bar{\kappa}_c(\bar{x}_c)$.
\end{corollary}

\begin{corollary}\label{cor:rpi}
  For each $c \in \mc{C}$, starting from $e_c(0) = 0$ the error
  dynamics
  $e_c^+ = A_{cc}e_c + B_{c}\left( \hat{\kappa}_c(\bar{e}_c;
    \bar{\mb{w}}_c) + \tilde{\kappa}_c(\hat{e}_c)\right) + w_c$ evolve
  in a robust positively invariant set
  $\mc{ER}_c \subseteq (\beta_{c}^x + \xi_c^x) \mbb{X}_c$.
\end{corollary}
The set $\mc{ER}_c$ is difficult to characterize, given its dependence
on the feasibility set of the secondary MPC controller,
$\bar{\mc{E}}^H_{c}(\bar{\mb{w}}_c)$, with the latter 
parameterized by $\bar{\mb{w}}_c$. Nevertheless, the basic
principle of tube-based robust MPC---that the trajectory of the
uncertain system is contained within a tube around the trajectory of
the nominal system---holds.

\section{Selection of the system partition}
\label{sec:switching_between_partitions}
In this section we consider the problem of choosing a suitable
partition for the system. The overall cost in~\eqref{eq:infine_cost}
measures the performance and practical costs of using different
partitions over time. However, the infinite-horizon combinatorial
optimization problem implied by minimizing~\eqref{eq:infine_cost} is
intractable. Our approach is therefore
to decouple the problems of regulation and partition selection: the
previous section presented a regulation algorithm for fixed
coalitions; in this section, we develop a partition selection
algorithm, using consensus optimization, to select the system
partition at a fixed state; in Section~\ref{sec:docp_implementation},
the partition selection and regulation algorithms are combined to
produce the overall approach, which varies the system partition in
time while regulating its states.

\subsection{Consensus optimization problem for partition selection}
\label{sec:consensus_partition}

The solution we propose to selecting the system partition is as
follows: each subsystem $i\in\mc{M}$ has an initial opinion on the
system partition, then an iteration process begins where subsystems
exchange information until a consensus on the system partition is
reached. With the system at a state $x$, we define the following
consensus optimization problem for subsystem $i \in \mc{M}$:
\begin{equation}
  \min \left\{ J_i (\mc{C}_{[i]}; \mc{C}_{[-i]},x)\colon\mc{C}_{[i]} \in\Pi_\mc{M}\right\}
\label{eq:consensus_problem}  
\end{equation}
% \mbb{C}(\mc{C}_{[i]}^-)
where $J_i(\mc{C}_{[i]}; \mc{C}_{[-i]},x) =  J_i^\ts{consensus} + \rho J_i^\ts{power}$ and
\begin{align*}
  J_i^\ts{consensus} &\triangleq \sum_{j\in\mc{M}_i}  w_{ij}(x_i,x_j)  \bigl| \delta_{ij}(\mc{C}_{[i]}) - \delta_{ij}(\mc{C}_{[j]}) \bigr| \\
                       J_i^\ts{power} &\triangleq \sum_{j \in \mc{M}_i} (w_{ij}(x_i,x_j) + \epsilon  ) \sigma_{ij} \bigl(1-\delta_{ij}(\mc{C}_{[i]})\bigr)  
\end{align*}
In this problem, the decision variable $\mc{C}_{[i]}$ is subsystem
$i$'s opinion on the system partition $\mc{C}$;
$\mc{C}_{[-i]} \triangleq \bigl\{ \mc{C}_{[j]} \bigr\}_{j \in
  \mc{M}_i}$ is the collection of neighbour's opinions, assumed fixed
at the point of solving this problem. The collection of all opinions,
written $(\mc{C}_{[1]},\dots,\mc{C}_{[M]})$---or equivalently
$(\mc{C}_{[i]},\mc{C}_{[-i]})$ for some $i$---is called a
\emph{profile}.

The objective function comprises two terms: the first term, as common
in consensus, penalizes differences between the opinion of subsystem
$i$ and the opinions of its neighbours, via the following indicator
function. For $i \in \mc{M}$,
\begin{equation*}
  \delta_{ij}(\mc{C}) = \delta_{ji}(\mc{C})=
  \begin{cases} 0 & \text{if} \ (i,j) \in c \ \text{for some} \ c \in \mc{C}, \\
      1 & \text{otherwise}.
      \end{cases}
    \end{equation*}
    Thus, $J^\ts{consensus}_i = 0$ if subsystem $i$ and its neighbours
    $j \in \mc{M}_i$ agree on being within, or \emph{not} being within, the
    same coalition.

    The second term in the objective penalizes a weighted function of
    \emph{link power} of coalitions in subsystem $i$'s opinion. The
    scalar $\sigma_{ij} = \sigma_{ji} > 0$ is the \emph{power} associated with the
    \emph{link} between subsystems $i$ and $j$ that being in a
    coalition together implies. In the simplest case,
    $\sigma_{ij} = 1$ so that
    $\sum_{j \in \mc{M}_i} \sigma_{ij}\bigl(1-\delta_{ij}(\mc{C}_{[i]})\bigr)$ merely
    counts the number of neighbours of subsystem $i$ contained within
    the same coalition. More sophisticated approaches have used
    game-theoretic measures with the aim of the link power accurately
    capturing the cost of using each link versus the benefit it brings
    to closed-loop performance~\citep{MAM+17a}.

    Both terms are weighted by a state-dependent function that
    measures the coupling strength between a pair of subsystems:
  \begin{equation*}
    w_{ij}(x_i,x_j) = w_{ji}(x_j,x_i)  = \frac{1}{2}\left(\| A_{ij} \| | x_j | + \| A_{ji} \| | x_i |\right).
  \end{equation*}

  The overall effect is as follows: the minimization of the consensus
  term in the objective promotes agreement on the system partition,
  with higher preference for putting more tightly coupled subsystems
  into the same coalition. The second term, penalizing weighted link
  power (where $\rho , \epsilon > 0$), is included as a
  regularization term. The absence of this may lead to pathological
  cases: in particular, a solution profile
  $(\mc{C}^e_{[1]},\ldots,\mc{C}^e_{[M]})$ for which
  $J^\ts{consensus}_i = 0$ does not necessarily imply a consensus, but
  something weaker in view of the state-dependent and possibly
  incomplete coupling structure. Some examples are illuminating in
  this regard.
\begin{example}[Dependence on coupling structure] For
	\begin{equation*}
		\begin{bmatrix} x_1 \\ x_2 \\ x_3 \end{bmatrix}^+ = \begin{bmatrix} 1 & a_{12} & 0 \\ a_{12} & 1 & a_{23} \\ 0 & a_{23} & 1 \end{bmatrix} \begin{bmatrix} x_1 \\ x_2 \\ x_3 \end{bmatrix} + Bu
	\end{equation*}
	with $x_i \neq 0$, $J^\ts{consensus}_i = 0$ implies
        consensus at any $\mc{C}^e \in \Pi_\mc{M}$ if $a_{12} \neq 0$
        and $a_{23} \neq 0$. If, however, $a_{23} = 0$ then the same
        implies $\mc{C}^e_{[1]} = \mc{C}^e_{[2]}$ but not necessarily
        $\mc{C}^e_{[2]} = \mc{C}^e_{[3]}$.
	\end{example}
\begin{example}[Dependence on state]\label{examp:2}
  For the same system, but now with $x_1 = x_2 = x_3 = 0$, $J_i^\ts{consensus} = 0$
for \emph{all} opinion profiles $\{ \mc{C}_{[1]},\dots,\mc{C}_{[M]} \}$.
\end{example}
\begin{example}[Regularization]\label{examp:3}
  For the scenario in Example~\ref{examp:2}, the profile
  $\mc{C}_{[1]} = \mc{C}_{[2]} = \mc{C}_{[3]} = \mc{M}$ (the
  decentralized partition) attains the minimum value of
  $J_i \bigl( \mc{C}_{[i]}, \mc{C}_{[-i]} \bigr)$, for all $i$, when all $\epsilon \sigma_{ij} > 0$.
\end{example}

\subsection{The consensus algorithm and its convergence}

As a prerequisite to the consensus-based algorithm for partition
selection, we establish useful properties of the \emph{game} between
subsystems that arises from the definition of the consensus
optimization problems. In particular, this forms a \emph{potential
  game}~\citep{MS96}, for which strong results apply to the
equilibrium actions of the game and convergence of algorithms to these
solutions.

\begin{definition}[Finite exact potential game
  (FEPG)] \label{def:FEPG} The game defined by a set of players
  $\mc{M} = \{1,\dots,M\}$ with finite action sets
  $\mc{A} = \{\mc{A}_i\}_{i\in \mc{M}}$ and objective functions
  $\{J_i: \mc{A}_i\times\mc{A}_{-i} \to \mbb{R} \}_{i \in \mc{M}}$, is
  a \emph{finite exact potential game} (FEPG) if there is a
  \emph{potential function} $\phi \colon \mc{A} \to \mbb{R}$ such
  that, for all $i \in \mc{M}$, $a_{-i} \in \mc{A}_{-i}$ and
  $a_i^\prime, a_i^{\prime\prime}\in\mc{A}_i$
  \begin{equation*}
  T_i(a_i^{\prime},a_{-i}) - T_i(a^{\prime\prime}_i,a_{-i}) = \phi(a_i^{\prime},a_{-i}) -
  \phi(a_i^{\prime\prime},a_{-i}).
\end{equation*}
\end{definition}
\begin{definition}[Nash equilibrium]
  An action profile $a^* = (a^*_{1},\dots,a^*_{M})$ is said to be a
  \emph{Nash equilibrium} of the game
  $\bigl(\mc{M},\{\mc{A}_i\}_i,\{T_i\}_i\bigr)$ if, for all
  $i \in \mc{M}$,
\begin{equation*}
  T_i(a^*_{i},a^*_{-i}) = \min_{a_{i} \in \mc{A}_i} T_i(a_{i},a^*_{-i}). 
  \end{equation*}
\end{definition}

The next result then follows immediately from the choice of objective
function in~\eqref{eq:consensus_problem}.

\begin{theorem}
  The game $(\mc{M},\{\Pi_\mc{M}\}_i,\{J_i\}_i)$
  is a finite exact potential game with potential function
  \begin{equation}
    \phi (\mc{C}_{[1]},\dots,\mc{C}_{[M]})  = \frac{1}{2} \sum_{i \in \mc{M}}  J_i^\ts{consensus} + \rho J_i^\ts{power} 
\label{eq:fepg_potential_function}
\end{equation}
Moreover, the game admits at least one Nash equilibrium; the set
of these equilibria coincides with the set of Nash equilibria for the
game $(\mc{M},\{\Pi_\mc{M}\}_i,\{\phi\}_i)$.
\label{thm:fepg}
\end{theorem}

We define the algorithm for playing this game by the recursion
\begin{equation}\label{eq:alg}
	\mc{C}_{[i]}(p+1) = \arg \min_{\mc{C}_{[i]}\in\mbb{C}_\Delta(\mc{C}_{[i]}(p))} \bigl\{ J_i (\mc{C}_{[i]}; \mc{C}_{[-i]}(p), x)\bigr\} 
\end{equation}
where $p$ is the iteration number and, without loss of generality, the
order of serial upating of subsystems' opinions is
$\{1,2,3,\dots,M\}$. The initial opinion of subsystem $i$ is
$\mc{C}_{[i]}(0) = \mc{C}_{[i]}^0$, and $\mc{C}_{[-i]}(p)$ denotes the
opinions of $i$'s neighbours at iteration $p$.

\begin{figure}[t] 
  \centering\footnotesize
% Minimal pgfplots version: 1.3
%
\definecolor{mc1}{rgb}{153,50,204}%
\definecolor{mc2}{rgb}{1.00000,1.00000,0.00000}%
\definecolor{mc3}{rgb}{0.00000,1.00000,1.00000}%

\newcommand{\myshorten}{.1em}
\tikzset{>=latex}

\begin{tikzpicture}[scale=0.3]
\node (L0) at (0,0) {$\{1,2,3,4\}$};
\node [below left=1.75em and 6em of L0] (L1)  {$\{1,2,3\},\{4\}$};
\node [below left=3.75em and 1.5em of L0] (L2)  {$\{1,2,4\},\{3\}$};
\node [below right=1.75em and 6em of L0] (L3) {$\{1,3,4\},\{2\}$};
\node [below right=3.75em and 1.5em of L0] (L4) {$\{1,3\},\{2,4\}$};
\node [below=3.75em of L0] (L5) {$\{2,3\},\{1,4\}$};
\node [above=0.5em of L5] (L6) {$\{1,2\},\{3,4\}$};
\node [below=0.5em of L5] (L7) {$\{2,3,4\},\{1\}$};

\node [below=5em of L1] (L8) {$\{1,2\},\{3\},\{4\}$};
\node [below=5em of L2] (L9) {$\{1,3\},\{2\},\{4\}$};
\node [below=5em of L3] (L10) {$\{3,4\},\{1\},\{2\}$};
\node [below=5em of L4] (L11) {$\{2,4\},\{1\},\{3\}$};
\node [below=3em of L5] (L12) {$\{1,4\},\{2\},\{3\}$};
\node [below=6em of L5] (L13) {$\{2,3\},\{1\},\{4\}$};

\node [below=1em of L13](L14)  {$\{1\},\{2\},\{3\},\{4\}$};

%\draw [<->,blue,  thick,shorten <=-\myshorten,shorten >=-\myshorten] (L0) to [out=210, in=90] (L2);
%\draw [<->,green,  thick,shorten <=-\myshorten,shorten >=-\myshorten] (L0) to [out=330, in=90] (L4);
\draw [<->,blue,  thick,shorten <=-\myshorten,shorten >=-\myshorten] (L0) to [out=360, in=90] (L3);
%\draw [<->,orange,  thick,shorten <=-\myshorten,shorten >=-\myshorten] (L0) to [out=240, in=180] (L5);
%\draw [<->,brown,  thick,shorten <=-\myshorten,shorten >=-\myshorten] (L0) to [out=270, in=90] (L6);

%\draw [<->,magenta,  thick,shorten <=-\myshorten,shorten >=-\myshorten] (L1) to [out=270, in=90] (L9);
%\draw [<->,magenta,  thick,shorten <=-\myshorten,shorten >=-\myshorten] (L1) to [out=270, in=150] (L13);

%\draw [<->,blue,  thick,shorten <=-\myshorten,shorten >=-\myshorten] (L2) to [out=270, in=90] (L8);
%\draw [<->,blue,  thick,shorten <=-\myshorten,shorten >=-\myshorten] (L2) to [out=270, in=90] (L11);
%\draw [<->,blue,  thick,shorten <=-\myshorten,shorten >=-\myshorten] (L2) to [out=270, in=120] (L12);

\draw [<->,blue,  thick,shorten <=-\myshorten,shorten >=-\myshorten] (L3) to [out=270, in=90] (L10);
%\draw [<->,yellow,  thick,shorten <=-\myshorten,shorten >=-\myshorten] (L3) to [out=270, in=30] (L12);
%\draw [<->,yellow,  thick,shorten <=-\myshorten,shorten >=-\myshorten] (L3) to [out=270, in=60] (L9);

%\draw [<->,green,  thick,shorten <=-\myshorten,shorten >=-\myshorten] (L4) to [out=270, in=90] (L11);
%\draw [<->,green,  thick,shorten <=-\myshorten,shorten >=-\myshorten] (L4) to [out=270, in=90] (L10);
%\draw [<->,green,  thick,shorten <=-\myshorten,shorten >=-\myshorten] (L4) to [out=270, in=30] (L13);

%\draw [<->,brown,  thick,shorten <=-\myshorten,shorten >=-\myshorten] (L6) to [out=225, in=90] (L8);
%\draw [<->,brown,  thick,shorten <=-\myshorten,shorten >=-\myshorten] (L6) to [out=315, in=90] (L10);

%\draw [<->,orange,  thick,shorten <=-\myshorten,shorten >=-\myshorten] (L5) to [out=216, in=160] (L13);
%\draw [<->,orange,  thick,shorten <=-\myshorten,shorten >=-\myshorten] (L5) to [out=324, in=20] (L12);

%\draw [<->,purple,  thick,shorten <=-\myshorten,shorten >=-\myshorten] (L7) to [out=270, in=90] (L9);

%\draw [<->,  thick,shorten <=-\myshorten,shorten >=-\myshorten] (L9) to [out=270, in=90] (L14);

\draw [<->,blue,  thick,shorten <=-\myshorten,shorten >=-\myshorten] (L10) to [out=270, in=0] (L14);

%\draw [<->,  thick,shorten <=-\myshorten,shorten >=-\myshorten] (L12) to [out=195, in=90] (L14);
%\draw [<->,  thick,shorten <=-\myshorten,shorten >=-\myshorten] (L13) to [out=270, in=90] (L14);

% chain examples
{\draw [<->,red,  thick,shorten <=-\myshorten,shorten >=-\myshorten] (L0) to [out=-180, in=90] (L1);
\draw [<->,red,  thick,shorten <=-\myshorten,shorten >=-\myshorten] (L1) to [out=270, in=90] (L8);
\draw [<->,red,  thick,shorten <=-\myshorten,shorten >=-\myshorten] (L8) to [out=270, in=180] (L14);\label{fig:chain1}}

\draw [<->,purple!70!black, thick,shorten <=-\myshorten,shorten >=-\myshorten] (L0) to [out=300, in=90] (L4);
\draw [<->,purple!70!black, thick,shorten <=-\myshorten,shorten >=-\myshorten] (L4) to [out=270, in=90] (L11);
\draw [<->,purple!70!black, thick,shorten <=-\myshorten,shorten >=-\myshorten] (L11) to [out=270, in=30] (L14);\label{fig:chain2}

\draw [<->,dashed,green!50!black, ,shorten <=-\myshorten,shorten >=-\myshorten] (L1) to [out=0, in=180] (L2);
\draw [<->,dashed,green!50!black,shorten <=-\myshorten,shorten >=-\myshorten] (L2) to [out=0, in=180] (L5);
\draw [<->,dashed,green!50!black, ,shorten <=-\myshorten,shorten >=-\myshorten] (L2) to [out=0, in=180] (L6);
\draw [<->,dashed,green!50!black, ,shorten <=-\myshorten,shorten >=-\myshorten] (L2) to [out=0, in=180] (L7);
\draw [<->,dashed,green!50!black,,shorten <=-\myshorten,shorten >=-\myshorten] (L5) to [out=0, in=180] (L4);
\draw [<->,dashed,green!50!black, ,shorten <=-\myshorten,shorten >=-\myshorten] (L6) to [out=0, in=180] (L4);
\draw [<->,dashed,green!50!black,,shorten <=-\myshorten,shorten >=-\myshorten] (L7) to [out=0, in=180] (L4);
\draw [<->,dashed,green!50!black,,shorten <=-\myshorten,shorten >=-\myshorten] (L4) to [out=0, in=180] (L3);\label{fig:antichain1}
\end{tikzpicture}%
%
\caption{Refinement relation defined over $\Pi_{\{1,2,3,4\}}$. Each path of different colour points towards the elements that are related to each other, through refinement. Solid lines represent elements of chains whereas dashed lines represent members of anti-chains.}
    \label{fig:partition_3set}
\end{figure}
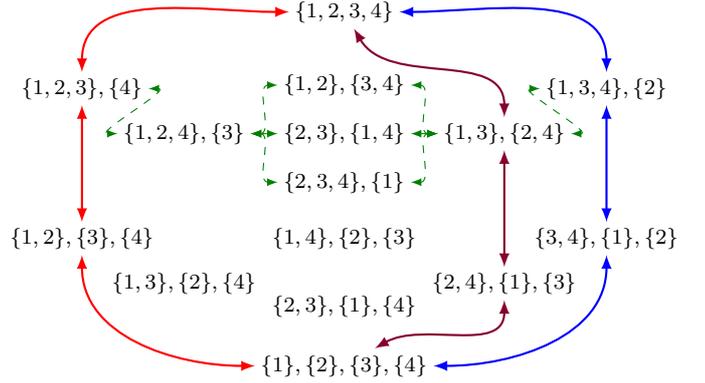

In this problem, subsystem $i$ seeks to determine an optimal opinion
$\mc{C}_{[i]}(p+1)$ from a \emph{subset} of the set of partitions
$\Pi_\mc{M}$.  The aim of this restriction is to reduce the
cardinality of the decision space, and hence complexity of the
optimization problem, since the full partition set $\Pi_\mc{M}$ grows
combinatorially with the number of subsystems. In particular, given
$i$'s previous opinion $\mc{C}_{[i]}(p)$, the domain of the problem is
$\mbb{C}_\Delta(\mc{C}_{[i]}(p)) \subset \Pi_\mc{M}$, which restricts
the choice of $\mc{C}_{[i]}$ to $\Delta \geq 1$ moves along
\emph{chains} of the partition set containing $\mc{C}_{[i]}(p)$. A
chain in the poset $\Pi_\mc{M}$ is any pair of elements that are
comparable under the \emph{refinement} order relation, denoted by
`$\preceq$', and defined as follows: given
$\mc{C}, \mc{D} \in \Pi_{\mc{M}}$, the partition
$\mc{D} \preceq \mc{C}$ ($\mc{D}$ \emph{refines} $\mc{C}$, or $\mc{C}$
\emph{coarsens} $\mc{D}$) if every member of $\mc{D}$ is contained in
some member of $\mc{C}$. For example, the partition
$\bigl\{\{1,2\},\{3,4\}\bigr\}$ refines
$\bigl\{\{1,2,3,4\}\bigr\}$. Because the ordering is partial, however,
not all pairs of partitions are comparable: an \emph{anti-chain} of
$\Pi_\mc{M}$ is a set of incomparable elements. The Hasse diagram in
Figure~\ref{fig:partition_3set} illustrates some chains and
anti-chains in the partition set for~$\mc{M} = \{1,2,3,4\}$.

Executing the algorithm defines a \emph{path} of opinions,
$ \mf{C} = \bigl\{ \bigl(\mc{C}_{[i]}(p), \mc{C}_{[-i]}(p)\bigr)
\bigr\}_{p,i}$, where $p=0,1,\dots$ and $i = (p+1) \bmod M$. The path
$\mf{C}$ is called an \emph{improvement path} if, for all $p \geq 1$,
$\phi \bigl(\mc{C}_{[i]}(p+1),\mc{C}_{[-i]}(p)\bigr) < \phi
\bigl(\mc{C}_{[i]}(p),\mc{C}_{[-i]}(p)\bigr) $. It is well known that
for an FEPG the path of best responses constitutes a finite
improvement path, and terminates in a Nash equilibrium; however,
because we have restricted the domain of the optimization
problem~\eqref{eq:alg} from $\Pi_\mc{M}$ to
$\mbb{C}_\Delta(\mc{C}_{[i]}(p))$, the algorithm is no longer a
straightforward application of an FEPG and these facts need to be
established separately.

\begin{theorem}[Finite improvement path]\label{thm:FIP}
  If $\Delta \geq 1$, then $\mf{C}$ is a finite improvement path,
  terminating in a Nash equilibrium
  $(\mc{C}_{[1]}^e,\dots,\mc{C}_{[M]}^e)$ that is a minimum of the
  potential function~\eqref{eq:fepg_potential_function}.
  \end{theorem}
 
\begin{remark}
  The restriction of the domain to lower the complexity of the
  optimization problem does not destroy the finite improvement
  property of the potential game; however, it will affect the rate of
  convergence to an equilibrium.
  \end{remark}

  The final result of gives some interesting implications about the
  nature of the attained equilibrium, depending on the relative
  weightings of the power term and consensus term in the overall cost.

  \begin{theorem}[When consensus implies equilibrium]\label{thm:NE}
    Suppose $(\mc{C}_{[1]}^e,\dots,\mc{C}_{[M]}^e)$ is such that
    $J^\ts{consensus}_i = 0$ for all $i \in \mc{M}$. If
    $\rho \sigma_{ij} < 1$ for all $i, j$ and $\epsilon \geq 0$ is
    sufficiently small, then $(\mc{C}_{[1]}^e,\dots,\mc{C}_{[M]}^e)$
    is a Nash equilibrium.
\end{theorem}

 \begin{remark}
   We note that $\epsilon = 0$ will always be sufficiently small, but
   such a choice will remove the regularizing behaviour of this
   additional term, as Examples~\ref{examp:2} and \ref{examp:3}
   show. It is interesting to note that Theorem~\ref{thm:NE} confirms
   that when all $x_i = 0$, any $\epsilon > 0$ and all
   $\sigma_{ij} > 0$ will result in a potential decrease from a point
   $(\mc{C}_{[1]}^e,\dots,\mc{C}_{[M]}^e)$ with
   $J^\ts{consensus}_i = 0$ until the decentralized partition is
   reached.
   \end{remark}

\section{Coalitional MPC with time-varying coalitions}
\label{sec:docp_implementation}
The final part of the development is to unite the coalitional
regulation scheme of Section~\ref{sec:dmpc_fixed_partition} with the
partition selection algorithm of
Section~\ref{sec:switching_between_partitions}, and study the
properties of the overall approach. We first note that the timescales
of the regulation and partition selection algorithms are not
coupled. Even though the regulation algorithm computes a new control
input to apply to the system at every time $k$, there is no assumption
or requirement on the rate of the partition selection algorithm: it
may, for example, propose a new partition at every sampling instance
or less frequently. In any case, significant challenges arise:
recursive feasibility is the most basic requirement for any MPC
controller, since guaranteeing the stability of the controlled system
relies on the continued feasibility of the underlying optimal control
problem. While recursive feasibility for each time-invariant coalition
is established by Proposition~\ref{prop:feas}, this does not continue
to hold for a time-varying system partition. Even assuming this can be
established, stability does not necessarily follow: the system of
time-varying coalitions is akin to a system that switches between
independently stable modes; it is well known that the stability of
such systems can be lost through excessively fast switching. These
aspects of the approach are discussed in detail in this section. Our
aim is to determine conditions under which a new system partition
proposed by the consensus algorithm is feasible to adopt and maintains
closed-loop stability.

\subsection{Recursive feasibility for time-varying coalitions}
\label{sec:feassec}

\subsubsection{Definitions}

The set
$\bar{\mc{X}}^N_\mc{C}$ is defined as the product of the individual
feasibility sets of the coalitions:
\begin{equation*}
\bar{\mc{X}}^N_\mc{C} \triangleq \prod_{c \in \mc{C}} \bar{\mc{X}}^N_c, 
\end{equation*}
with a similar definition for $\mc{ER}_\mc{C}$. Note that the
condition $x \in \bar{\mc{X}}^N_\mc{C}$ is equivalent to the condition
$x_c \in \bar{\mc{X}}^N_c$ for all $c \in \mc{C}$, but utilizes, for
convenience, a more compact notation.

We also introduce notions of \emph{feasibility} and \emph{strong
  feasibility} with respect to a partition. In the context of adopting
a new system partition, strong feasibility has the advantage of
permitting the simple initialization of coalitional controller states
that is proposed in Algorithm~\ref{alg:nedmpc}, avoiding an otherwise
iterative and coupled design process.

\begin{definition}[Feasible and strongly feasible partition]
  A partition is said to be \emph{feasible} at a state $x$ if
  $x \in \bar{\mc{X}}^N_\mc{C} \oplus \mc{ER}_\mc{C}$, and
  \emph{strongly feasible} at a state $x$ if
  $x \in \bar{\mc{X}}^N_\mc{C}$.
\end{definition}

\subsubsection{Basic results on how partition coarsening and refinement affect feasibility}

Any change in system partition proposed by the selection algorithm is
either a coarsening or a refinement. The next set of results presented
are therefore relevant and useful, having fundamental implications for
allowing a change in system partition online. The first result is a
consequence of the fact that, with coarsening, the disturbance set
that each coalition sees diminishes, leading to a smaller RCI set. A
similar relation holds with partition refinement: the disturbance set
grows leading to a larger RCI set.
\begin{theorem}[Nesting of RCI sets]\label{thm:nest}
  Suppose that
  Assumptions~\ref{assump:controllability}--\ref{assump:constraint_sat}
  hold.  If $\mc{C} \succeq \mc{D}$, then
  ${\mc{R}}_\mc{C} \subseteq {\mc{R}}_\mc{D}$.
\end{theorem}

This, in turn, implies less restriction of the constraints in
the primary MPC problems under coarsening, and more restriction with refinement.
\begin{corollary}[Nesting of nominal feasibility regions]\label{lem:3}
  If $\mc{C} \succeq \mc{D}$, then
  $\bar{\mc{X}}^i_\mc{C} \supseteq \bar{\mc{X}}^i_\mc{D}$ for
  $i = 0\dots N$.
  \label{lem:nesting}
\end{corollary}

This might seem to suggest that feasibility is trivially
maintained with partition coarsening. The reality is, however, not so
simple, owing to the following corollary and the fact that
$x \in \mc{X}_\mc{C} \oplus \mc{ER}_\mc{C}$.

\begin{corollary}[Counter-nesting of error feasibility regions]
  If $\mc{C} \succeq \mc{D}$, then
  $\mc{ER}_\mc{C} \subseteq \mc{ER}_\mc{D}$.
\end{corollary}

\begin{proposition}[Feasibility is not necessarily maintained with coarsening]\label{prop:nofeas}
  Suppose $\mc{D} \in \Pi_\mc{M}$ is feasible at $x$. Then
  $\mc{C} \succeq \mc{D}$ is not necessarily feasible at $x$.
\end{proposition}

That is, even though
$\bar{\mc{X}}^N_\mc{C} \supseteq \bar{\mc{X}}^N_\mc{D}$ when
$\mc{C} \succeq \mc{D}$, there is no clear relation between the sets
$\mc{X}_\mc{D} \oplus \mc{ER}_\mc{D}$ and
$\mc{X}_\mc{C} \oplus \mc{ER}_\mc{C}$; the numerical examples
illustrate (and hence prove) this. On the other hand, \emph{strong}
feasibility is guaranteed under the same assumptions, as a consequence
of Corollary~\ref{lem:nesting}.
    
\begin{proposition}[Strong feasibility is maintained with coarsening]
  Suppose $\mc{D} \in \Pi_\mc{M}$ is strongly feasible at a state
  $x$. Then $\mc{C} \succeq \mc{D}$ is strongly feasible at $x$.
  \label{prop:str_coarse}
\end{proposition}
  
The situation is more challenging in the case of partition refinement,
since a counterpart to Proposition~\ref{prop:str_coarse} for a
movement from $\mc{C}$ to $\mc{D} \preceq \mc{C}$ does not hold.

\begin{proposition}[Strong feasibility does not imply feasibility after refinement]
  Suppose $\mc{C} \in \Pi_\mc{M}$ is strongly feasible at a state
  $x$. Then $\mc{D} \preceq \mc{C}$ is not necessarily feasible at $x$.
  \label{prop:str_refine}
\end{proposition}

This raises at least two questions: firstly, \emph{when} is the
hypothesis of Proposition~\ref{prop:str_coarse}, for partition
coarsening, met? Even though the initial state
$x(0)\in\bar{\mc{X}}^N_\mc{C}$ implies that the nominal state
$\bar{x}(k) \in \bar{\mc{X}}^N_\mc{C}$ for all $k \in \mbb{I}_{>0}$, it
\emph{does not} imply that the true state
$x(k) \in \bar{\mc{X}}^N_\mc{C}$. Secondly, when is strong feasibility
achieved under partition refinement? In the next subsection, we
present and discuss answers to these questions.

\subsubsection{Schemes for feasible partition switching}

We outline three schemes for enabling feasible switching between
partitions over time. Our intention is not to develop any scheme into
a comprehensive proposal, but to explore the range of options and
illustrate the comparative ease or difficulty of implementing each.

\paragraph{A quest for feasibility by design}

With the system in a partition $\mc{C}$ that is feasible at a state
$x$, the nominal state $\bar{x} \in \bar{\mc{X}}^N_\mc{C}$. Under
Algorithm~\ref{alg:nedmpc}, the successor nominal state
$\bar{x}^+ \in \bar{\mc{X}}^{N-1}_\mc{C}$ and the true state
$x^+ \in \bar{\mc{X}}^{N-1}_\mc{C} \oplus \mc{ER}_\mc{C} \subset
\bar{\mc{X}}^{N}_\mc{C} \oplus \mc{ER}_\mc{C}$. This motivates the
following proposition concerning switching between partitions,
offering two possible ways to ensure strong feasibility.

\begin{proposition}
  Suppose partition $\mc{C} \in \Pi_\mc{M}$ is feasible at a state $x$. Partition $\mc{D} \in \Pi_\mc{M}$ is strongly feasible for the successor state $x^+$ if $\bar{\mc{X}}^{N-1}_\mc{C} \oplus \mc{ER}_\mc{C} \subseteq \bar{\mc{X}}^{N}_\mc{D}$,
  which is satisfied if
  \begin{equation*}
    \bar{\mc{X}}^{N-1}_\mc{C} \oplus  \prod_{c \in \mc{C}} (\beta_c^x + \xi_c^x) \mbb{X}_c  \subseteq \bar{\mc{X}}^{N}_\mc{D}.
    \end{equation*}
  \end{proposition}
  
  With respect to the \emph{usefulness} of this result, the first
  inclusion is not straightforward to verify or enforce, in view of
  the difficulty of characterizing $\mc{ER}_\mc{C}$. The second
  inclusion offers a more practical---since the scalars $\beta^x_c$ and
  $\xi^x_c$ are generated from the design procedure
  in~\ref{sec:design}---yet more conservative way to guarantee strong
  feasibility. However, the condition is still problematic to impose,
  and perhaps impossible to meet, as a design constraint because of
  the fundamental relations governing the relations between sets under
  refinement and coarsening. For example, for the condition to be met
  under the refinement $\mc{D} \preceq \mc{C}$, it is necessary that
  $\bar{\mc{X}}^{N-1}_\mc{C} \subset
  \interior{\bar{\mc{X}}^N_{\mc{D}}}$ even though
  $\bar{\mc{X}}^N_{\mc{C}} \supseteq \bar{\mc{X}}^N_{\mc{D}}$;
  possibilility of satisfaction would be highly problem specific and,
  even if possible, would require careful design. Even under
  coarsening, for which
  $\bar{\mc{X}}^N_{\mc{C}} \subseteq \bar{\mc{X}}^N_{\mc{D}}$ already,
  the condition is not trivially met and relies on weak coupling for
  satisfaction---the size of the coalitional disturbance set
  $\mbb{W}_c = \bigoplus_{d \in \mc{M}_c} A_{cd} \mbb{X}_d$ must be
  sufficiently small.  More constructive alternatives are therefore
  discussed next.

  \paragraph{Use of a feasibility dwell time}

  An attractive option, well established in the switched systems
  literature, and more recently in the context of MPC for
  switched systems~\citep{MMA12, ZZB16, HT19}, is the use of a
  \emph{dwell time} to ensure the state lies within the
  feasibility region for the new partition at the moment
  of switching. The next result, which follows directly from the
stability of each coalition in the time-invariant setting (Theorem~\ref{thm:stab}), enables this.

\begin{proposition}[Feasibility becomes and remains strong
  feasibility]\label{prop:feasbecomesstrong}
  Suppose the system is in a partition $\mc{C} \in \Pi_\mc{M}$ that is
  feasible at $x$, and controlled by Algorithm~\ref{alg:nedmpc}. The same partition $\mc{C}$ becomes, and remains, strongly feasible a finite number of timesteps thereafter. Moreover, if $\mc{ER}_\mc{C} \subset \bar{\mc{X}}_\mc{C}$, then this happens exponentially fast.  
\end{proposition}

The hypothesis $\mc{ER}_\mc{C} \subset \bar{\mc{X}}_\mc{C}$ is
satisfied if the constraint scaling factors follow
$\beta^x_c + \xi_c^x < \alpha_c^x$ for all $c \in \mc{C}$; note that
this is, again, a weak coupling requirement.

Once strong feasibility is established for all subsequent times, a
similar result establishes that a switch from partition $\mc{C}$ to
partition $\mc{D}$ is, if the coupling is sufficiently weak, possible
after a finite number of steps.

\begin{proposition}[Strong feasibility dwell time]\label{prop:feastime}
  Suppose the system is in a partition $\mc{C} \in \Pi_\mc{M}$ that is
  feasible at $x$. A partition $\mc{D} \neq \mc{C}$ becomes strongly
  feasible a finite number of time steps thereafter. Moreover, if
  $\mc{R}_\mc{C} \subset {\mc{X}}^N_\mc{D}$, then this happens exponentially fast.
\end{proposition}

\paragraph{It is not necessary to implement a candidate partition}

A key observation is that, unlike in the case of switched systems
where mode-to-mode switches are not necessarily something that can be
controlled, the choice of system partition at each time is a
controllable degree of freedom. It follows, then, that if the system
partition is $\mc{C}$ and, subsequently, a new partition $\mc{D}$ is
selected by the decision-making process, it is not \emph{necessary} to
adopt the new partition. Indeed, if the new partition $\mc{D}$ is not
(strongly) feasible, then Proposition~\ref{prop:feas} already ensures
that the current partition $\mc{C}$ is.

% This motivates the use of a simpler approach than the dwell-time-based
% one, of checking the strong feasibility of a partition immediately
% after it is selected. To this end, there are two options: explicit or implicit 
% checking via a set-membership test or convex
% optimization respectively. For the former, note that it is possible to obtain an
% explicit characterization of the set $\bar{\mc{X}}^N_c$, for each
% $c \in \mc{C}$, by performing $N$ iterations of the backwards
% reachability operator; the test of strong feasibility of a partition
% $\mc{C}$ amounts then to the verification that
% $x_c \in \bar{\mc{X}}^N_c$ for each $c \in \mc{C}$.

% Alternatively, feasibility may be checked by attempting to solve the
% QP $\bar{\mbb{P}}_c(\bar{x}_c)$ or, more simply, by solving the
% associated linear programming (LP) problem obtained by replacing the
% quadratic cost of $\bar{\mbb{P}}_c(\bar{x}_c)$ with a linear one; in
% the latter case, an infeasibility certificate is easy to obtain.

\subsection{Closed-loop stability for time-varying coalitions}

Our final development is to consider the impact of changing the system
partition on closed-loop stability. We recall the following result, concerning the multiple Lyapunov-like function approach to stability for switched systems.

\begin{lemma}[{\citet[Lemma 1]{Zhang2013}}]
  Consider the switched system $x(k+1) = f_{\sigma(k)}\left(x(k)\right)$, where
  $f_\sigma(\cdot)$ is globally Lipschitz continuous with
  $f_\sigma(0) = 0$, and $\sigma \colon \mbb{I}_{\geq 0} \to \Sigma$ is the switching signal that takes values in a finite set $\Sigma$.
  If there exists a family of continuous positive-definite functions $V_m : \Rset^n \to \Rset_{\geq 0}$, $\sigma(k) = m \in \Sigma$, satisfying for all $k \in \mbb{I}_{\geq 0}$ and all $m \in \Sigma$
    \begin{gather*}
    \alpha_1\left( | x(k) |\right) \leq V_m \left( x(k) \right) \leq \alpha_2 \left( | x(k) | \right)\\
    V_m\left( x(k) \right) \leq \mu V_{\sigma(0)}\left( x(0) \right)
    \end{gather*}
    for two $\mc{K}$-functions $\alpha_1(\cdot)$ and $\alpha_2(\cdot)$ and a positive scalar $\mu > 0$, then the origin is locally stable.
\label{lem:stable_switching}
\end{lemma}

Establishing closed-loop stability of the time-varying coalitional system then amounts to showing that the value functions of the coalitional MPC controllers satisfy the conditions of this lemma. In the following result, $\kappa_\mc{C}(\cdot)$ refers to the collection of control laws when the system is partitioned into $\mc{C} = \{1,\dots,C\} \in \Pi_\mc{M}$; more precisely, $\kappa_\mc{C}(x) = \left( \kappa_c(x_c) \right)_{c \in \mc{C}}$, where $\kappa_c(x_c) = \bar{\kappa}_c(\bar{x}_c) + \hat{\kappa}_c(\bar{e}_c; \bar{\mb{w}}_c) + \tilde{\kappa}_c(\hat{e}_c)$  is the three-term control policy for coalition $c$ defined in~\eqref{eq:control_law_nested}. The \emph{switching signal} $\sigma \colon \mbb{I}_{\geq 0} \mapsto \Pi_\mc{M}$ is implicitly defined by the partition selection algorithm; at time $k$, the algorithm has the system partitioned into a set of coalitions $\sigma(k) = \mc{C}(k) \in \Pi_\mc{M}$.

\begin{theorem}[Local stability]\label{thm:final}
  Suppose that
  Assumptions~\ref{assump:controllability}--\ref{assump:constraint_sat}
  hold, that $x(0) \in \bar{\mc{X}}^N_{\mc{C}(0)}$ and that the
  switching signal $\sigma(\cdot)$ is such that, at each moment of switching, strong feasibility of the next partition is attained. Then the origin is locally stable for ${x}^+=A{x} + B\kappa_\sigma(x)$.
\end{theorem}

Finally, the stronger result of asymptotic stability is obtained under the further assumption that the partition switching algorithm settles (in time, rather with iterations) to a time-invariant set of coalitions.

\begin{assumption}\label{assump:conv}
  The signal $\sigma(k) = \bar{\mc{C}} \in \Pi_\mc{M}$ for $k \geq k_f$ and some finite $k_f \geq 0$ such that $x(k_f) \in \bar{\mc{X}}^N_{\bar{\mc{C}}}$.
  \end{assumption}

\begin{theorem}[Asymptotic stability]\label{thm:conv}
  Under the same hypotheses as Theorem~\ref{thm:final}, plus Assumptions~\ref{assump:decent} and \ref{assump:conv}, the origin is asymptotically stable for ${x}^+=A{x} + B\kappa_\sigma(x)$. The region of attraction is $\bar{\mc{X}}^N_{\mc{C}(0)}$.
\end{theorem}
 %
% {\color{blue}
% \begin{remark}
% 	The existence of a dwell time
% \end{remark}}
%
\section{Illustrative examples}
\label{sec:simulations}
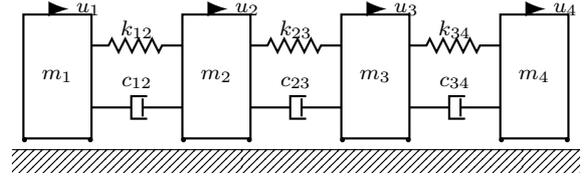
\begin{figure}
  \centering\footnotesize
  \begin{tikzpicture}[every node/.style={draw,outer sep=0pt,thick},scale = 0.25]
\tikzstyle{spring}=[thick,decorate,decoration={zigzag,pre length=0.75em,post length=0.75em,segment length=5}]
\tikzstyle{damper}=[thick,decoration={markings,  
  mark connection node=dmp,
  mark=at position 0.5 with 
  {
    \node (dmp) [thick,inner sep=0pt,transform shape,rotate=-90,minimum width=1em,minimum height=0.5em,draw=none] {};
    \draw [thick] ($(dmp.north east)+(0.2em,0)$) -- (dmp.south east) -- (dmp.south west) -- ($(dmp.north west)+(0.2em,0)$);
    \draw [thick] ($(dmp.north)+(0,-0.4em)$) -- ($(dmp.north)+(0,0.4em)$);
  }
}, decorate]
\tikzstyle{ground}=[fill,pattern color=black,pattern=north east lines,draw=none,minimum width=0.75cm,minimum height=0.3cm]

\node (M1) [minimum width=3em, minimum height=5.5em] {$m_1$};
\node (M2) [minimum width=3em, minimum height=5.5em, right=4em of M1] {$m_2$};
\node (M3) [minimum width=3em, minimum height=5.5em, right=4em of M2] {$m_3$};
\node (M4) [minimum width=3em, minimum height=5.5em, right=4em of M3] {$m_4$};

\node (ground) [ground,anchor=north,yshift=-0.5em,minimum width=25em] at ($(M1.south west)!0.5!(M4.south east)$) {};

\draw (ground.north east) -- (ground.north west);
\draw [thick] (M1.south west) ++ (0.2em,-0.25em) circle (0.25em)  (M1.south east) ++ (-0.2em,-0.25em) circle (0.25em);
\draw [thick] (M2.south west) ++ (0.2em,-0.25em) circle (0.25em)  (M2.south east) ++ (-0.2em,-0.25em) circle (0.25em);
\draw [thick] (M3.south west) ++ (0.2em,-0.25em) circle (0.25em)  (M3.south east) ++ (-0.2em,-0.25em) circle (0.25em);
\draw [thick] (M4.south west) ++ (0.2em,-0.25em) circle (0.25em)  (M4.south east) ++ (-0.2em,-0.25em) circle (0.25em);

\draw [spring] ($(M1.north east)!0.5!(M1.east)$) -- ($(M2.north west)!0.5!(M2.west)$) node[pos=0.5,above,draw=none]{$k_{12}$};
\draw [damper] ($(M1.south east)!0.5!(M1.east)$) -- ($(M2.south west)!0.5!(M2.west)$) node[pos=0.5,above=0.5em,draw=none]{$c_{12}$};

\draw [spring] ($(M2.north east)!0.5!(M2.east)$) -- ($(M3.north west)!0.5!(M3.west)$) node[pos=0.5,above,draw=none]{$k_{23}$};
\draw [damper] ($(M2.south east)!0.5!(M2.east)$) -- ($(M3.south west)!0.5!(M3.west)$) node[pos=0.5,above=0.5em,draw=none]{$c_{23}$};

\draw [spring] ($(M3.north east)!0.5!(M3.east)$) -- ($(M4.north west)!0.5!(M4.west)$) node[pos=0.5,above,draw=none]{$k_{34}$};
\draw [damper] ($(M3.south east)!0.5!(M3.east)$) -- ($(M4.south west)!0.5!(M4.west)$) node[pos=0.5,above=0.5em,draw=none]{$c_{34}$};

\draw [-latex,ultra thick] (M1.north) ++ (0,1em) -- +(2em,0em) node[pos=1,draw=none,right]{$u_1$};
\draw [-latex,ultra thick] (M2.north) ++ (0,1em) -- +(2em,0em)node[pos=1,draw=none,right]{$u_2$};
\draw [-latex,ultra thick] (M3.north) ++ (0,1em) -- +(2em,0em)node[pos=1,draw=none,right]{$u_3$};
\draw [-latex,ultra thick] (M4.north) ++ (0,1em) -- +(2em,0em)node[pos=1,draw=none,right]{$u_4$};

\end{tikzpicture}
  \caption{Coupled mass--spring--damper system}
  \label{fig:carts}%
  \end{figure}

\begin{table}[b]
\centering\footnotesize
\caption{Closed-loop performance costs and practical costs---where $\sigma_{ij} = 1$ for all $i,j$---for different partitions of the system.}
\begin{tabular}{ccrr}
	\toprule
	&Partition &$V^\infty(x,\mb{u})$ & $(1/2)\sum_{i} J_i^{\ts{power}} $\\
	\midrule                                                         
	$\mc{C}^1$ & $\{\{1,2,3,4\}\}$ & $73.1327$ &$1.2$\\
	$\mc{C}^2$ & $\{\{1,2,3\},\{4\}\}$ & $73.6061$ & $0.6$\\
	$\mc{C}^3$ & $\{\{1,2,4\},\{3\}\}$ & $73.8282$ & $0.6$\\
	$\mc{C}^4$ & $\{\{1,2\},\{4,3\}\}$ & $73.8376$ & $0.4$\\
	$\mc{C}^5$ & $\{\{1,2\},\{4\},\{3\}\}$ & $73.8270$ & $0.2$\\
	$\mc{C}^6$ & $\{1,3,4\},\{2\}$ & $73.3287$ & $0.6$\\
	$\mc{C}^7$ & $\{\{1,3\}\{2,4\}\}$ & --- & $0.4$\\
	$\mc{C}^8$ & $\{\{1,3\},\{2\},\{4\}\}$ & --- & $0.2$\\
	$\mc{C}^9$ & $\{\{1,4\},\{2,3\}\}$ & $73.2006$ & $0.4$\\
	$\mc{C}^{10}$ & $\{\{1\},\{2,3,4\}\}$ & $73.2008$ & $0.6$\\
	$\mc{C}^{11}$ & $\{\{1\},\{2,3\},\{4\}\}$ & $73.1890$ & $0.2$\\
	$\mc{C}^{12}$ & $\{\{1,4\},\{2\},\{3\}\}$ & $73.2027$ & $0.2$\\
	$\mc{C}^{13}$ & $\{\{1\},\{2,4\},\{3\}\}$ & $73.1910$ & $0.2$\\
	$\mc{C}^{14}$ & $\{\{1\},\{2\},\{3,4\}\}$ & $73.2016$ & $0.2$\\
	$\mc{C}^{15}$ & $\{\{1\},\{2\},\{3\},\{4\}\} $& $73.1910$ & $0.0$\\
	\bottomrule
\end{tabular}
\label{table:coal_costs}
\end{table}
 \begin{table}[b]
 	\centering\footnotesize
 	 	\caption{Cost function values for different algorithms.}
 		\begin{tabular}{lrrr}
 			\toprule
 			 & $V^\infty(x,\mb{u})$ & $J^\infty(x,\mc{C})$ & $J(\mc{C}(k))$\\
 				\midrule    
 			$\mc{C}^\ts{cen}$ & 176.86 & 120.00 & 1.20\\
 			$\mc{C}^\ts{dec}$ & 178.19 &  27.30 & 0.00\\
 			$\mc{C}(k)$ & 177.67 & 27.65 & 0.15\\
 				\bottomrule
 	\end{tabular}
 	\label{table:costs}
 \end{table}
We illustrate and explore the results via an example system of a
planar chain of four coupled mass--spring--dampers, shown in Figure~\ref{fig:carts}.
Each mass
corresponds to a subsystem, and the state of mass $i$ comprises its
position (relative to some datum) and velocity, $x_i = (r_i,v_i)$. The continuous-time dynamics are
\begin{align*}
\begin{split}
  \begin{bmatrix}
    \dot{r}_i \\ \dot{v}_i
  \end{bmatrix}
  &=
  \underbrace{\begin{bmatrix}
    0 & 1\\
    -\frac{1}{m_i}\sum_{j \in \mc{M}_i} k_{ij} & -\frac{1}{m_i}\sum_{j \in \mc{M}_i} c_{ij} 
    \end{bmatrix}}_{A_{ii}}
  \begin{bmatrix}
    {r}_i \\ {v}_i
  \end{bmatrix}
  \\ &\quad  +
  \begin{bmatrix}  
    0 \\ 100
  \end{bmatrix}
  u_i + w_i\end{split}\\
  w_i &= \sum_{j \in \mc{M}_i}
  \underbrace{\begin{bmatrix}
    0 & 0\\
    \frac{1}{m_i}\sum_{j \in \mc{M}_i} k_{ij} & \frac{1}{m_i}\sum_{j \in \mc{M}_i} c_{ij} 
    \end{bmatrix}}_{A_{ij}}
  \begin{bmatrix}
    {r}_j \\ {v}_j
  \end{bmatrix}.%
\end{align*}%
Here, $u_i$ is the control input (acceleration) to mass $i$. The disturbance
$w_i$ arises via the coupling between masses: mass 1 ($m_1=\SI{3}{\kilogram}$) is coupled to mass 2 ($m_2=\SI{2}{\kilogram}$) via a spring (stiffness
$k_{12}=\SI{0.5}{\newton\per\metre}$) and damper
($c_{12}=\SI{0.2}{\newton\per\metre\per\second}$). Likewise, mass 3
($m_3=\SI{3}{\kilogram}$) is coupled to mass 4 ($m_4=\SI{6}{\kilogram}$) via $k_{34}=\SI{1}{\newton\per\metre}$ and
$c_{34}=\SI{0.3}{\newton\per\metre\per\second}$. Moreover, masses 2 and 3 are also coupled, via
$k_{23}=\SI{0.75}{\newton\per\metre}$ and
$c_{23}=\SI{0.25}{\newton\per\metre\per\second}$.

The system is subject to constraints on the states,
$\mbb{X}_i= \left\{(r_i,v_i): -2\leq r_i\leq 2,\ -8\leq v_i\leq
  8\right\}$, and control inputs,
$\mbb{U}_i= \left\{u_i : -4\leq u_i\leq 4 \right\}$. The initial
conditions are $x_1(0) = (1.8,0)$, $x_2(0) = (-0.5,0)$,
$x_3(0) = (1,0)$, and $x_4(0) = (-1,0)$. The parameters for the MPC
controllers are a horizon $N = H - 1 = 25$ and matrices
$Q_{i} = I$, $R_{i} = 1$.
\subsection{Performance and practical costs with time-invariant partitions}
The initial exploration is the performance of the system under
different, albeit static, partitions. For $\mc{M}=\{1,2,3,4\}$, there are $15$ possible partitions in the partition set
$\Pi_\mc{M} = \{\mc{C}^1,\mc{C}^2,\ldots,\mc{C}^{14},\mc{C}^{15}\}$,
where $\mc{C}^1$ corresponds to the centralized partition
$\mc{C}^\ts{cen}$ and $\mc{C}^{15}$ corresponds to the decentralized
partition $\mc{C}^\ts{dec}$. For each partition, the primary and
secondary MPC controllers were designed for each coalition using the
procedure in~\ref{sec:design}. The design succeeded for all partitions
except $\mc{C}^7 = \{\{1,3\},\{2,4\}\}$ and
$\mc{C}^8 = \{\{1,3\},\{2\},\{4\}\}$. Inspection revealed
that $\mc{C}^7$ and $\mc{C}^8$ place subsystems with no physical
coupling into coalitions; consequently, the interactions between
coalitions are too strong in order for the design procedure to
succeed, and therefore there do not exist suitable scaling factors to build the optimal control problems in these cases.

The closed-loop state trajectories for mass $1$ are shown in
Figure~\ref{fig:coal_performance_comparison}. The whole-system
closed-loop costs are in Table~\ref{table:coal_costs}.  The
centralized partition achieves the lowest regulation cost but has the
highest practical cost at each step. The opposite outcome is observed
for the decentalized partition. The performance of $\mc{C}^3$,
$\mc{C}^4$, and $\mc{C}^5$ exhibit the worst regulation performance,
as a consequence of these coalitions leading to the highest constraint
tightening margins, but with lower (higher) practical cost than for
the centralized (decentralized). On the other hand, some surprising
results emerge: performance does not necessarily always deteriorate
with refinement (compare $\mc{C}^{11}$ with $\mc{C}^2$), which
suggests a complicated relationship between partition and closed-loop
performance that demands further research.
\subsection{Time-varying partitions}
To increase coupling strength, we change the initial conditions to $x_1(0) = (1.0,-7)$, $x_2(0) = (-0.51,4)$, $x_3(0) = (-1.71,0)$, and $x_4(0) = (1.8,-4)$. The aim is to investigate the performance of the overall scheme, including the consensus-based partition selection algorithm. To this end, the partition opinions for each subsystem are initialized, at iteration $p=0$ and time $k = 0$, as $\mc{C}_{[1]}(0) = \mc{C}^2$, $\mc{C}_{[2]}(0) = \mc{C}^2$, $\mc{C}_{[3]}(0) = \mc{C}^{10}$, and $ \mc{C}_{[4]}(0) = \mc{C}^{14}$.
%
%\begin{align*}
%\mc{C}_{[1]}(0) &= \mc{C}^2 & \mc{C}_{[2]}(0) &= \mc{C}^2\\
%\mc{C}_{[3]}(0) &= \mc{C}^{10} & \mc{C}_{[4]}(0) &= \mc{C}^{14}\\
%\end{align*}
%

The execution of the partition selection algorithm at time $k=0$ is
illustrated in Figure~\ref{fig:consensus_convergence_it0}; the
outcome, after eight iterations, is a consensus among subsystems to
adopt the centralized partition, $\mc{C}^1$, for the initial time
$k=0$. Figure~\ref{fig:consensus_convergence} shows the system partition
selected and employed at each subsequent time step during the
simulation. Following the use of initial centralized partition, the
subsystems agree on the partition $\mc{C}^{10}$---which groups
subsystem $2$, $3$, and $4$ together, and subsystem $1$
separately---between times $k=1$ and $k=7$. At time $k = 8$, the
partition changes to $\mc{C}^{14}$, which groups only subsystems $3$
and $4$. At $k=19$ the subsystems agree to disband into the decentralized partition $\mc{C}^{15}$, as the effect of the dynamic
coupling weakens, \ie $w_{ij}(x_i,x_j)\to 0$.
\begin{figure}[t]
	\centering\footnotesize
	\input{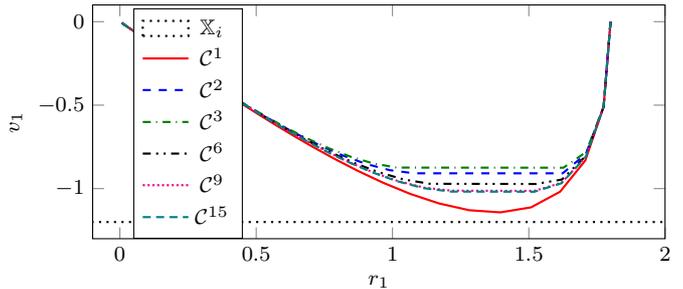}%
	\caption{Closed-loop state trajectories for different system partitions. $\mc{C}^4$ and $\mc{C}^5$ gave responses visually indistinguishable from that of $\mc{C}^3$; likewise, $\mc{C}^{10}$--$\mc{C}^{14}$ gave responses in between those of $\mc{C}^9$ and $\mc{C}^{15}$.}
    \label{fig:coal_performance_comparison}
\end{figure}
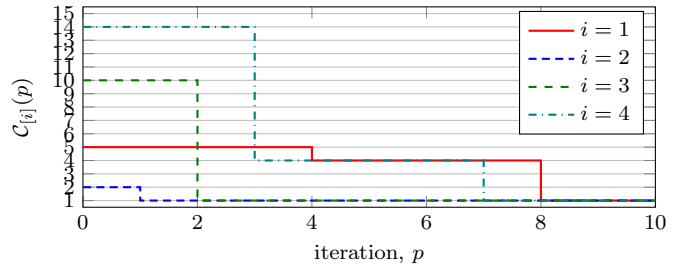
\begin{figure}[t]
	\centering\footnotesize
\definecolor{cyan}{rgb}{0.00000,1.00000,1.00000}
%color= cyan!50!black
\begin{tikzpicture}

\begin{axis}[%
enlargelimits=false,
width=0.85\linewidth,
height=0.3\linewidth,
scale only axis,
xmin=0,
xmax=10,
xlabel={iteration, $p$},
ymin=0.5,
ymax=15.5,
ytick = {1,2,3,4,5,6,7,8,9,10,11,12,13,14,15},
ymajorgrids,
ylabel={$\mc{C}_{[i]}(p)$},
]
\addplot [color=red,solid,thick]
  table[row sep=crcr]{%
0	5\\
1	5\\
2	5\\
3	5\\
4	5\\
4	4\\
5	4\\
6	4\\
7	4\\
8	4\\
8	1\\
9	1\\
10 1\\
};\label{fig:c10}
\addlegendentry{$i = 1$}

\addplot [densely dashed, color=blue,thick]
  table[row sep=crcr]{%
 0	2\\
1	2\\
1	1\\
2	1\\
3	1\\
4	1\\
5	1\\
6	1\\
7	1\\
8	1\\
9	1\\
10 1\\
};\label{fig:c20}
\addlegendentry{$ i =2$}

\addplot [dashed, color=green!50!black,thick]
  table[row sep=crcr]{%
0	10\\
1	10\\
1	10\\
2	10\\
2	1\\
3	1\\
4	1\\
5	1\\
6	1\\
7	1\\
8	1\\
9	1\\
10 1\\
};\label{fig:c30}
\addlegendentry{$i = 3$}

\addplot [dashdotted, color=cyan!50!black,thick]
  table[row sep=crcr]{%
0	14\\
1	14\\
2	14\\
3	14\\
3	4\\
4	4\\
5	4\\
6	4\\
7	4\\
7	1\\
8	1\\
9	1\\
10 1\\
};\label{fig:c40}
\addlegendentry{$i = 4$}

\end{axis}
\end{tikzpicture}%
%
\caption{Consensus algorithm iterations at the initial time step $k = 0$.}
    \label{fig:consensus_convergence_it0}
\end{figure}
\begin{figure}[t]
	\centering\footnotesize
% Minimal pgfplots version: 1.3
%
\begin{tikzpicture}

\begin{axis}[%
enlargelimits=false,
width=0.85\linewidth,
height=0.3\linewidth,
scale only axis,
xmin=0,
xmax=30,
xlabel={time step, $k$},
ymin=0.5,
ymax=15.5,
ytick = {1,2,3,4,5,6,7,8,9,10,11,12,13,14,15},
ymajorgrids,
ylabel={$\mc{C}(k)$},
]

\addplot [color=red,solid,thick]
  table[row sep=crcr]{%
0	1\\
1	1\\
1	10\\
2	10\\
3	10\\
4	10\\
5	10\\
6	10\\
7	10\\
8	10\\
8	14\\
9	14\\
10	14\\
11	14\\
12	14\\
13	14\\
14	14\\
15	14\\
16	14\\
17	14\\
18	14\\
19 14\\
19	15\\
20	15\\
21	15\\
22	15\\
23	15\\
24	15\\
25	15\\
26	15\\
27	15\\
28	15\\
29	15\\
30	15\\
};\label{fig:L1}
%\addlegendentry{$\mc{C}(k)$}

%\addplot [color=green!50!black,solid,line width=1.5pt]
%  table[row sep=crcr]{%
%0	2\\
%1	2\\
%1	15\\
%2	15\\
%3	15\\
%4	15\\
%4	15\\
%4	15\\
%5	15\\
%6	15\\
%7	15\\
%8	15\\
%9	15\\
%10	15\\
%11	15\\
%12	15\\
%13	15\\
%14	15\\
%15	15\\
%16	15\\
%17	15\\
%18	15\\
%19	15\\
%20	15\\
%};\label{fig:L2}
%\addlegendentry{$\Lambda_2(k)$}
%\addplot [color=blue!50!black,solid,line width=1.5pt]
%  table[row sep=crcr]{%
%0	10\\
%1	10\\
%2	10\\
%2	14\\
%3	14\\
%4	14\\
%5	14\\
%6	14\\
%6	15\\
%7	15\\
%8	15\\
%9	15\\
%10	15\\
%11	15\\
%12	15\\
%13	15\\
%14	15\\
%15	15\\
%16	15\\
%17	15\\
%18	15\\
%19	15\\
%20	15\\
%};\label{fig:L3}
%\addlegendentry{$\Lambda_3(k)$}
%\addplot [color=pink!50!black,solid,line width=1.5pt]
%  table[row sep=crcr]{%
%0	14\\
%1	14\\
%2	14\\
%3	14\\
%4	14\\
%5	14\\
%6	14\\
%7	14\\
%7	15\\
%8	15\\
%9	15\\
%10	15\\
%11	15\\
%12	15\\
%13	15\\
%14	15\\
%15	15\\
%16	15\\
%17	15\\
%18	15\\
%19	15\\
%20	15\\
%};\label{fig:L4}
%\addlegendentry{$\Lambda_4(k)$}
%\addplot [color=black,dashed,line width=3.5pt]
%  table[row sep=crcr]{%
%0	2\\
%1	2\\
%1	15\\
%2	15\\
%2	14\\
%3	14\\
%4	14\\
%4	15\\
%5	15\\
%6	15\\
%7	15\\
%8	15\\
%9	15\\
%10	15\\
%11	15\\
%12	15\\
%13	15\\
%14	15\\
%15	15\\
%16	15\\
%17	15\\
%18	15\\
%19	15\\
%20	15\\
%};\label{fig:sig}
%\addlegendentry{$\sigma(k)$}
\end{axis}
\end{tikzpicture}%
%
\caption{Partition selected at each time step.}
    \label{fig:consensus_convergence}
\end{figure}
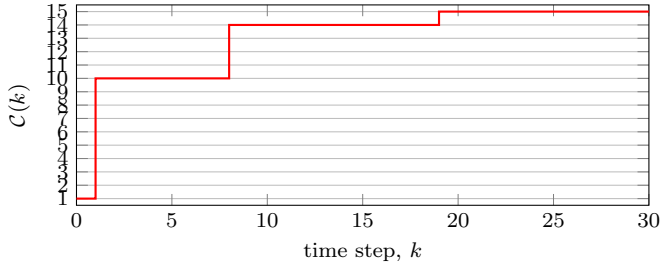
\begin{figure}[t]
	\centering\footnotesize
% Minimal pgfplots version: 1.3
%
\begin{tikzpicture}

\begin{axis}[%
enlargelimits=false,
width=0.85\linewidth,
height=0.3\linewidth,
scale only axis,
xmin=-2,
xmax=0.1,
xmajorgrids,
ymin=-4.2,
ymax=2,
ymajorgrids,
ylabel={$v_{4}$},
xlabel={$r_{4}$},
legend style={at={(axis cs:-0.25,-1)},anchor=north},
]

\addplot [color=blue,solid,thick]
  table[row sep=crcr]{%
-1.8	-4\\
-1.96387999122905	0.718922757598607\\
-1.84763838753174	1.60484810233838\\
-1.68439263646431	1.65956784388386\\
-1.52416611229821	1.54461335427214\\
-1.37669086295556	1.40459865325284\\
-1.24293727535107	1.27021354469329\\
-1.12205794014432	1.1471399413146\\
-1.01290787968369	1.03565097758249\\
-0.914369716750234	0.934922501503167\\
-0.825416303473589	0.843974458669008\\
-0.745116344339149	0.761870086557409\\
-0.672628246433071	0.687752278473015\\
-0.607192092842317	0.620844780597959\\
-0.548121851190824	0.56044629878122\\
-0.494798219840755	0.505923641022605\\
-0.446662143519261	0.45670518806466\\
-0.403208953926387	0.412274924447034\\
-0.363983079082685	0.372167033767398\\
-0.328573267092183	0.33596101610754\\
-0.296608274497537	0.303277279658681\\
-0.267752974078498	0.273773161037933\\
-0.241704841299008	0.247139332728456\\
-0.218190782568541	0.223096560012612\\
-0.196964272064882	0.201392773404216\\
-0.177802767100146	0.181800425888003\\
-0.160505374932032	0.164114107259558\\
-0.144890746558544	0.148148390553276\\
-0.130795175414134	0.133735887980372\\
-0.118070881033505	0.120725495994093\\
-0.106584459688533	0.108980811083469\\
-0.0962154857543807	0.0983786996861113\\
-0.0868552491411577	0.0888080072264957\\
-0.0784056155540288	0.0801683927451896\\
-0.0707779976324484	0.0723692769006997\\
-0.0638924261816754	0.0653288923147259\\
-0.0576767117591279	0.0589734263041178\\
-0.0520656878254384	0.0532362470119844\\
-0.0470005275252125	0.0480572048243028\\
-0.0424281269344564	0.0433820017479345\\
-0.0383005483084872	0.039161622138487\\
-0.0345745174932	0.0353518188096348\\
-0.0312109702304319	0.0319126491360002\\
-0.0281746426007816	0.0288080562861809\\
-0.0254337013099864	0.0260054911952786\\
-0.0229594099426993	0.0234755713136486\\
-0.0207258276845879	0.0211917725542299\\
-0.0187095373540894	0.0191301512085389\\
-0.0168893998924491	0.0172690929159629\\
-0.0152463327380548	0.0155890860545284\\
-0.0137631097614976	0.0140725171771436\\
-0.0124241806638335	0.0127034863488124\\
-0.0112155079445808	0.0114676404485278\\
-0.010124419730186	0.0103520226885743\\
-0.00913947691998061	0.00934493677296163\\
-0.00825035325676453	0.00843582427135459\\
-0.00744772706464483	0.00761515392250316\\
-0.00672318351908884	0.00687432170680945\\
-0.00606912642457325	0.00620556064014725\\
-0.00547869857488592	0.00560185934351809\\
};\label{fig:xcen}
\addlegendentry{$\mc{C}^{\ts{cen}}$}

\addplot [color=green!50!black,solid,thick,dashed]
  table[row sep=crcr]{%
-1.8	-4\\
-1.95131752676031	0.969927967741344\\
-1.85298885334003	0.99635033813053\\
-1.72484484991433	1.56569928146329\\
-1.5685221766125	1.56032442234016\\
-1.41844427500194	1.44091622771986\\
-1.28099172037265	1.30786392480021\\
-1.15647711713262	1.18218699319005\\
-1.04398142328493	1.06750995664174\\
-0.94241014473849	0.963719960559272\\
-0.850716876814133	0.869968817697019\\
-0.767944150748597	0.785326312393966\\
-0.693224818357917	0.708916454432232\\
-0.625775477281561	0.639940485830978\\
-0.564888810672564	0.577675602355805\\
-0.509926291819018	0.521468938343635\\
-0.460311514717753	0.470731064978\\
-0.415524155077507	0.424929884845238\\
-0.375094514827712	0.383585068435048\\
-0.3385985965374	0.346263020468142\\
-0.305653656885807	0.312572333459092\\
-0.275914192728319	0.28215968308047\\
-0.249068317701074	0.254706122777353\\
-0.224834491373066	0.229923738882039\\
-0.202958566645137	0.207552631173995\\
-0.183211124431432	0.187358187224301\\
-0.165385067671295	0.169128621941305\\
-0.14929344943928	0.152672756513758\\
-0.134767512375954	0.137818013458674\\
-0.121654918878366	0.124408606747404\\
-0.109818153489632	0.112303908028924\\
-0.0991330807329732	0.10137697181675\\
-0.0894876432657544	0.0915132041722891\\
-0.080780686700644	0.0826091609228755\\
-0.0729208987694286	0.0745714628108956\\
-0.0658258517041517	0.0673158161968943\\
-0.0594211377927265	0.0607661290464313\\
-0.0536395890433228	0.0548537129296902\\
-0.0484205727738961	0.0495165626652247\\
-0.0437093557394933	0.0446987060526646\\
-0.0394565301287442	0.0403496168754325\\
-0.0356174954097668	0.0364236850172806\\
-0.0321519905914624	0.0328797381350543\\
-0.0290236719948733	0.0296806098723014\\
-0.0261997321065582	0.0267927500844239\\
-0.023650555516787	0.0241858729885352\\
-0.0213494083342737	0.0218326395473258\\
-0.0192721578202408	0.019708370756738\\
-0.0173970193015351	0.0177907888298005\\
-0.0157043277085986	0.0160597835631151\\
-0.0141763313423439	0.0144972014350686\\
-0.012797005707097	0.0130866552249271\\
-0.0115518854572221	0.0118133521549761\\
-0.0104279126950141	0.0106639387545114\\
-0.00941330002889251	0.00962636081824491\\
-0.0084974069557602	0.0086897369902588\\
-0.0076706282711164	0.00784424464812327\\
-0.00692429333663413	0.00708101689040361\\
-0.00625057514880149	0.00639204954674421\\
-0.00564240825499946	0.00577011723609362\\
};\label{fig:xdec}
\addlegendentry{$\mc{C}^{\ts{dec}}$}
\addplot [color=red,dashdotted,thick]
  table[row sep=crcr]{%
-1.8	-4\\
-1.96387999122905	0.718922757598607\\
-1.89152980082409	0.727872613625673\\
-1.77750699152648	1.5515803726509\\
-1.62001195369503	1.59784433036738\\
-1.4658091596883	1.48587650861562\\
-1.32395588338878	1.35090745013203\\
-1.19531823259409	1.22159599466498\\
-1.07906636296725	1.10321712167645\\
-0.974095813221372	0.995991639897729\\
-0.87933122216391	0.899117671831928\\
-0.793784571995494	0.811650588705735\\
-0.716560157295451	0.732688992121466\\
-0.646848573134114	0.661408446247797\\
-0.583918973778442	0.597062356338355\\
-0.527111574254403	0.538976239259436\\
-0.475830768915971	0.486541115276208\\
-0.429538894652857	0.439207225011617\\
-0.38775059539192	0.396478287833033\\
-0.350027733653873	0.357906303426485\\
-0.315973270070047	0.323117362312383\\
-0.285230128457489	0.291686264012745\\
-0.257477777431671	0.263307314958597\\
-0.232425586734185	0.237688258059718\\
-0.209810916251145	0.214561604627027\\
-0.189396616626766	0.193685077985932\\
-0.170968598738873	0.174839794701425\\
-0.154333600947583	0.157828128702871\\
-0.139317165030591	0.142471673924942\\
-0.125761807168996	0.12860938105255\\
-0.113525366477439	0.116095870236223\\
-0.102479514128351	0.104799906804721\\
-0.0925084075163084	0.0946030250299436\\
-0.0835074753660416	0.0853982857410111\\
-0.0753823210469286	0.0770891548081761\\
-0.0680477325933898	0.0695884907463654\\
-0.0614267890489992	0.0628176308227712\\
-0.0554500537621688	0.056705566085248\\
-0.050054846173133	0.0511881966590395\\
-0.0451845844551343	0.0462076595024884\\
-0.0407881921157762	0.0417117215711178\\
-0.0368195623353004	0.0376532320262617\\
-0.0332370744240618	0.033989627742831\\
-0.0300031573280692	0.0306824869308236\\
-0.0270838956048643	0.0276971261888525\\
-0.0244486737374343	0.025002236763735\\
-0.0220698550559225	0.0225695562016136\\
-0.0199224918998329	0.0203735719470975\\
-0.0179840639810684	0.0183912537817589\\
-0.0162342422039042	0.0166018122959465\\
-0.0146546754649659	0.0149864808611336\\
-0.0132287981972805	0.0135283188160549\\
-0.0119416566400422	0.012212033802328\\
-0.0107797520120977	0.011023821387009\\
-0.00973089894443439	0.00995122028933052\\
-0.00878409768700189	0.00898298169385255\\
-0.00792941874963966	0.0081089512791143\\
-0.00715789876728397	0.00731996272501437\\
-0.00646144649734797	0.00660774158147982\\
-0.00583275796342924	0.0059648184908576\\
};\label{fig:xcoal}
\addlegendentry{$\mc{C}(k)$}
\end{axis}
\end{tikzpicture}%
%
\caption{State trajectory for subsystem $4$ under control by the centralized partition, decentralized partition and the proposed time-varying scheme.} 
    \label{fig:x_time}
\end{figure}
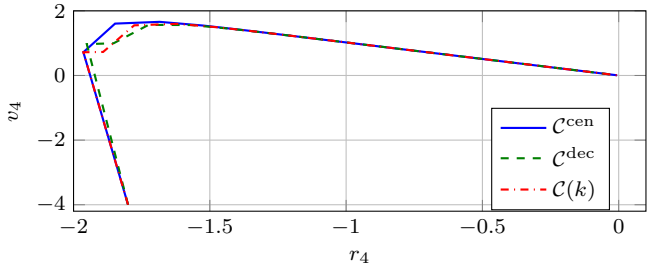
\begin{figure}[t!]
	\centering
\begin{tikzpicture}

\begin{axis}[%
enlargelimits=false,
width=0.85\linewidth,
height=0.45\linewidth,
scale only axis,
xmin=-1.1,
xmax=1.1,
%xmajorgrids,
xlabel={$x_1$},
ymin=-1.1,
ymax= 1.1,
%ymajorgrids,
ylabel={$x_2$},
zmin=-1.1,
zmax=1.1,
%zmajorgrids,
zlabel={$x_3$},
legend style={at={(1,1)},anchor=north east},
view={45}{55},
name={one}
]
\addplot3[area legend,solid,thick,draw=blue]
table[row sep=crcr] {%
x	y	z\\
0.972222222222222	-0.833333333333333	-0.99537037037037\\
0.972222222222222	-0.833333333333334	0.671296296296296\\
0.694444444444444	0.833333333333333	0.717592592592593\\
0.694444444444444	0.833333333333333	-0.949074074074074\\
}--cycle;\label{fig:s11}
\addlegendentry{$\mc{X}^1_{\mc{C}^\ts{cen}}$}

\addplot3[area legend,solid,thick,draw=blue,forget plot]
table[row sep=crcr] {%
x	y	z\\
-0.694444444444444	-0.833333333333333	-0.717592592592593\\
-0.694444444444444	-0.833333333333334	0.949074074074074\\
0.972222222222222	-0.833333333333334	0.671296296296296\\
0.972222222222222	-0.833333333333333	-0.99537037037037\\
}--cycle;

\addplot3[area legend,solid,thick,draw=blue,forget plot]
table[row sep=crcr] {%
x	y	z\\
0.972222222222222	-0.833333333333334	0.671296296296296\\
-0.694444444444444	-0.833333333333334	0.949074074074074\\
-0.972222222222222	0.833333333333333	0.99537037037037\\
0.694444444444444	0.833333333333333	0.717592592592593\\
}--cycle;

\addplot3[area legend,solid,thick,draw=blue,forget plot]
table[row sep=crcr] {%
x	y	z\\
0.694444444444444	0.833333333333333	-0.949074074074074\\
0.694444444444444	0.833333333333333	0.717592592592593\\
-0.972222222222222	0.833333333333333	0.99537037037037\\
-0.972222222222223	0.833333333333333	-0.671296296296297\\
}--cycle;

\addplot3[area legend,solid,thick,draw=blue,forget plot]
table[row sep=crcr] {%
x	y	z\\
-0.694444444444444	-0.833333333333333	-0.717592592592593\\
0.972222222222222	-0.833333333333333	-0.99537037037037\\
0.694444444444444	0.833333333333333	-0.949074074074074\\
-0.972222222222223	0.833333333333333	-0.671296296296297\\
}--cycle; 

\addplot3[area legend,solid,thick,draw=blue,forget plot]
table[row sep=crcr] {%
x	y	z\\
-0.694444444444444	-0.833333333333334	0.949074074074074\\
-0.694444444444444	-0.833333333333333	-0.717592592592593\\
-0.972222222222223	0.833333333333333	-0.671296296296297\\
-0.972222222222222	0.833333333333333	0.99537037037037\\
}--cycle;

\addplot3[area legend,dashed,thick,draw=green!50!black]
table[row sep=crcr] {%
x	y	z\\
0.972222222222223	-0.833333333333333	-0.807413333333333\\
0.972222222222223	-0.833333333333333	0.807413333333333\\
0.694444444444445	0.833333333333333	0.807413333333333\\
0.694444444444445	0.833333333333333	-0.807413333333333\\
}--cycle;\label{fig:s12}
\addlegendentry{$\mc{X}^1_{\mc{C}^2}$}

\addplot3[area legend,dashed,thick,draw=green!50!black,forget plot]
table[row sep=crcr] {%
x	y	z\\
-0.694444444444444	-0.833333333333333	-0.807413333333333\\
-0.694444444444444	-0.833333333333333	0.807413333333333\\
0.972222222222223	-0.833333333333333	0.807413333333333\\
0.972222222222223	-0.833333333333333	-0.807413333333333\\
}--cycle;

\addplot3[area legend,dashed,thick,draw=green!50!black,forget plot]
table[row sep=crcr] {%
x	y	z\\
0.972222222222223	-0.833333333333333	0.807413333333333\\
-0.694444444444444	-0.833333333333333	0.807413333333333\\
-0.972222222222221	0.833333333333333	0.807413333333333\\
0.694444444444445	0.833333333333333	0.807413333333333\\
}--cycle;

\addplot3[area legend,dashed,thick,draw=green!50!black,forget plot]
table[row sep=crcr] {%
x	y	z\\
-0.694444444444444	-0.833333333333333	0.807413333333333\\
-0.694444444444444	-0.833333333333333	-0.807413333333333\\
-0.972222222222221	0.833333333333333	-0.807413333333333\\
-0.972222222222221	0.833333333333333	0.807413333333333\\
}--cycle;
\addplot3[area legend,dashed,thick,draw=green!50!black,forget plot]
table[row sep=crcr] {%
x	y	z\\
0.694444444444445	0.833333333333333	-0.807413333333333\\
0.694444444444445	0.833333333333333	0.807413333333333\\
-0.972222222222221	0.833333333333333	0.807413333333333\\
-0.972222222222221	0.833333333333333	-0.807413333333333\\
}--cycle;

\addplot3[area legend,dashed,thick,draw=green!50!black,forget plot]
table[row sep=crcr] {%
x	y	z\\
-0.694444444444444	-0.833333333333333	-0.807413333333333\\
0.972222222222223	-0.833333333333333	-0.807413333333333\\
0.694444444444445	0.833333333333333	-0.807413333333333\\
-0.972222222222221	0.833333333333333	-0.807413333333333\\
}--cycle;

\addplot3[area legend,dotted,thick,draw=red]
table[row sep=crcr] {%
x	y	z\\
0.807413333333333	-0.833333333333333	-0.807413333333333\\
0.807413333333333	-0.833333333333333	0.807413333333333\\
0.807413333333333	0.833333333333333	0.807413333333333\\
0.807413333333333	0.833333333333333	-0.807413333333333\\
}--cycle;\label{fig:s13}
\addlegendentry{$\mc{X}^1_{\mc{C}^\ts{dec}}$}

\addplot3[area legend,dotted,thick,draw=red,forget plot]
table[row sep=crcr] {%
x	y	z\\
-0.807413333333333	-0.833333333333333	-0.807413333333333\\
-0.807413333333333	-0.833333333333333	0.807413333333333\\
0.807413333333333	-0.833333333333333	0.807413333333333\\
0.807413333333333	-0.833333333333333	-0.807413333333333\\
}--cycle;

\addplot3[area legend,dotted,thick,draw=red,forget plot]
table[row sep=crcr] {%
x	y	z\\
0.807413333333333	-0.833333333333333	0.807413333333333\\
-0.807413333333333	-0.833333333333333	0.807413333333333\\
-0.807413333333333	0.833333333333333	0.807413333333333\\
0.807413333333333	0.833333333333333	0.807413333333333\\
}--cycle;

\addplot3[area legend,dotted,thick,draw=red,forget plot]
table[row sep=crcr] {% 
x	y	z\\
-0.807413333333333	-0.833333333333333	0.807413333333333\\
-0.807413333333333	-0.833333333333333	-0.807413333333333\\
-0.807413333333333	0.833333333333333	-0.807413333333333\\
-0.807413333333333	0.833333333333333	0.807413333333333\\
}--cycle;

\addplot3[area legend,dotted,thick,draw=red,forget plot]
table[row sep=crcr] {%
x	y	z\\
0.807413333333333	0.833333333333333	-0.807413333333333\\
0.807413333333333	0.833333333333333	0.807413333333333\\
-0.807413333333333	0.833333333333333	0.807413333333333\\
-0.807413333333333	0.833333333333333	-0.807413333333333\\
}--cycle;

\addplot3[area legend,dotted,thick,draw=red,forget plot]
table[row sep=crcr] {%
x	y	z\\
-0.807413333333333	-0.833333333333333	-0.807413333333333\\
0.807413333333333	-0.833333333333333	-0.807413333333333\\
0.807413333333333	0.833333333333333	-0.807413333333333\\
-0.807413333333333	0.833333333333333	-0.807413333333333\\
}--cycle;

%% \addplot3[color=green,dashed,line width=1.5pt,mark = *]
%% table[row sep=crcr] {%
%% x	y	z\\
%%    0.972200000000000  -0.833300000000000   0.807400000000000\\
%%   -0.000000000000000                   0   0.085253669514310\\
%%   -0.000000000000000                   0  -0.000000000000004\\
%%   -0.000000000000000                   0   0.018724996675938\\
%%   -0.000000000000000                   0   0.005115944763333\\
%%   -0.000000000000000                   0   0.001397751248210\\
%%   -0.000000000000000                   0   0.000381886166596\\
%%   -0.000000000000000                   0   0.000104336908447\\
%%   -0.000000000000000                   0   0.000028506375488\\
%%   -0.000000000000000                   0   0.000007788360373\\
%%   -0.000000000000000                   0   0.000002127894424\\
%% };\label{fig:s14}
%% \addlegendentry{$\mb{x}(k)$}
\end{axis}

\end{tikzpicture}%
%
\caption{Feasibility regions for the partitions $\mc{C}^\ts{cen}$, $\mc{C}^2$ and $\mc{C}^\ts{dec}$.} %: (top) plotted in $\mbb{R}^3$, and (bottom) the projection onto $x_1$--$x_2$ space.
    \label{fig:feas_regions}
\end{figure}
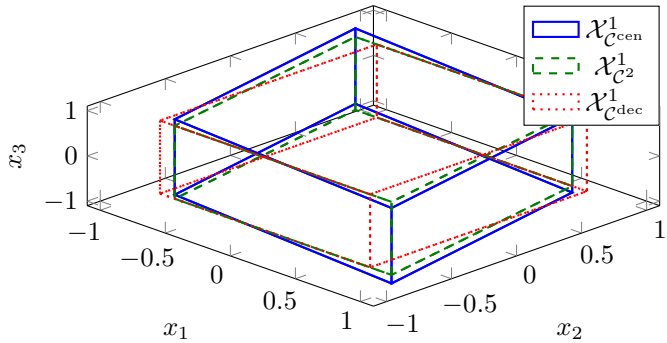

Finally, we compare the performance of the time-varying scheme against
that of a fixed centralized partition and a fixed decentralized
partition. Figure~\ref{fig:x_time} shows the closed-loop state
trajectory for subsystem 4, for which differences can be seen during
the transient. Table~\ref{table:costs} summarizes the closed-loop
costs. The first column gives the closed-loop performance costs, the
second column shows the summation (over the simulation) of the
consensus optimization cost \ie
$J^\infty(x,\mc{C}) = (1/2)\sum_{k}
\sum_{i\in\mc{M}}J_i(\mc{C}_{[i](k)};\mc{C}_{[-i]}(k),x(k)) $, and the
third column gives the value of
$(1/2)\sum_{i \in \mc{M}} J_i^\ts{power}$, where all
$\sigma_{ij} = 1$, averaged across the simulation. We observe that the
proposed scheme balances the trade-off between closed-loop performance
and practical cost.
\subsection{Feasibility regions under different partitions}
A smaller-scale example is now considered to illustrate the feasibility results
reported in Section~\ref{sec:feassec}. Let
\begin{equation}
  x_i^+=0.6x_i + u_i + w_i \ \text{for} \ i = 1, 2, 3,
\label{eq:coal_example}  
\end{equation}
where $w_1 = 0.1x_2$, $w_2 = 0$ and $w_3 = 0.1 x_1$. The subsystems
$i = 1,2,3$ are subject to constraints $|x_i|\leq 2$ and
$|u_i|\leq 0.5$.

There are five possible partitions of $\mc{M}$. We consider the
chain $\mbb{C} = \{\mc{C}^\ts{cen},\mc{C}^2,\mc{C}^\ts{dec}\}$, where
$\mc{C}^2=\{\{1,2\},\{3\}\}$. The one-step feasibility regions for the
\emph{true} coalition dynamics, \ie~the product set of
${\mc{X}}^1_{c} \triangleq \bar{\mc{X}}^1_c \oplus
{\mc{ER}}^1_c$ over $c \in \mc{C}$, for
each $\mc{C} \in \mbb{C}$, is displayed in
Figure~\ref{fig:feas_regions}.

Illustrating Proposition~\ref{prop:nofeas} and the subsequent
discussion, there is no nesting of the sets
$\mc{X}^1_{\mc{C}^\ts{cen}}$, $\mc{X}^1_{\mc{C}^2}$, and
$\mc{X}^1_{\mc{C}^\ts{dec}}$. Indeed, there exist states
$x\in\mc{X}_{\mc{C}^2}^1$ such that
$x\notin\mc{X}_{\mc{C}^\ts{cen}}^1$ and/or
$x\notin\mc{X}_{\mc{C}^\ts{dec}}^1$; for example, the state
$x_1 = 0.9722$, $x_{2} = -0.8333$, $x_{3} = 0.8074$. Applying the
proposed algorithm from this state, we find that the system may only
be feasibly controlled if partitioned as $\mc{C}^2$. At later times,
once the state has been steered into
$\bar{\mc{X}}_{\mc{C}^\ts{dec}}^1$, the algorithm switches to the
decentralized partition.

\section{Conclusions}
\label{sec:conclusions}
A coalitional MPC approach for dynamically coupled subsystems was
presented. The approach controls a constrained system of subsystems in a
way that balances the performance degradation of decentralized control
with the practical cost of centralized control, by allowing subsystem
controllers to form coalitions over time, thus reconfiguring the
partition of the system dynamics. For regulation, the approach employs a robust distributed model predictive control with implicit,
rather than explicit, reliance on robust invariant
sets. Re-partitioning of the system into different coalitions is
achieved via a consensus-based algorithm. Recursive feasibility and
stability of the overall time-varying coalitional control scheme are
established under dwell-time assumptions. An interesting consequence
of the developments is the finding that a coalitional controller may
reach states that are otherwise infeasible for a centralized
controller.

\section*{References}
\bibliography{\myreferences}

\appendix

\section{Controller design}
\label{sec:design}

The design of the primary and secondary MPC controllers for each
coalition, via the selection of the scaling factors that restrict the
constraints in the optimal control problems, is based on the theory of
optimized robust control invariance~\citep{RKM+07}; therefore, we
begin with an overview of the main concepts.

\subsection{Optimized Robust Control Invariance}\label{sec:rci}
The optimized RCI approach~\citep{RKM+07}
proposed a novel characterization of an RCI set for a system $x^+ =
Ax+Bu+w$ and constraint set $(\mbb{X},\mbb{U},\mbb{W})$ as
\begin{equation*}
\mc{R}_h(\mb{M}_h) = \bigoplus\limits_{l=0}^{h-1}D_l(\mb{M}_h)\mbb{W} \ \text{with}\ \mu(\mc{R}_h(\mb{M}_h)) = \bigoplus\limits_{l=0}^{h-1} M_l \mbb{W}.
\label{eqn:rci_set}
\end{equation*}
The set $\mu(\mc{R}_h(\mb{M}_h))$ is the set of invariance-inducing
control actions, defined as $\mu(\mc{R}_h)\triangleq\{\mu(x):x \in
\mc{R}_h\} =\{u\in\mbb{U}: x^+\in\mc{R}_h, \forall w\in\mbb{W}\}$. The
matrices $D_l(\mb{M}_h), l = 0\dots h$ are
\begin{equation*}
 D_l(\mb{M}_h) = \begin{dcases} I & l = 0 \\
   A^l + \sum_{j=0}^{l-1} A^{l-1-j}BM_j & l \geq 1,
   \end{dcases}
  \end{equation*}
with $M_j \in \mbb{R}^{m\times n}$, and $\mb{M}_h \triangleq
(M_0,M_1,\dots,M_{h-1})$, such that $D_h(\mb{M}_h) = 0$; the latter is
ensured by setting $h$ greater than or equal to the controllability
index of $(A,B)$. The set of matrices that satisfy these conditions is
given by $\mbb{M}_h\triangleq \{\mb{M}_h : D_h(\mb{M}_h)=0\}$.
Constraint satisfaction is guaranteed if $\mc{R}_h(\mb{M}_h) \subseteq
\eta \mbb{X}$ and $\mu(\mc{R}_h(\mb{M}_h)) \subseteq \theta \mbb{U}$,
with $(\eta,\theta) \in [0,1] \times [0,1]$. The linear programming (LP) problem to
compute these matrices is
\begin{equation}\label{eq:optRCI}
\min \{ \delta : \gamma \in \Gamma\},
\end{equation}
where $\gamma=(\mb{M}_h,\eta,\theta,\delta)$, and the set
$\Gamma=\{\gamma : \mb{M}_h \in\mbb{M}_h,
\mc{R}_h(\mb{M}_h)\subseteq\eta\mbb{X},
\mu(\mc{R}_h(\mb{M}_h))\subseteq\theta\mbb{U},
(\eta,\theta)\in[0,1]\times
   [0,1],\ q_{\eta}\eta+q_{\theta}\theta\leq\delta\}$; $q_{\eta}$ and
   $q_{\theta}$ are weights to express a preference for the relative
   contraction of state and input constraint sets. Feasibility of this
   problem is linked to the existence of an RCI set:
   if~\eqref{eq:optRCI} is feasible, then $\mc{R}_h(\mb{M}_h)$ exists
   and satisfies the RCI properties~\citep{RKM+07}.

\subsection{Design algorithm}

The RCI LP problem is useful in the current context because solving it
provides an invariance-inducing robust control law---a suitable
candidate for the third term in the overall control law---plus scaling
constants that outer-bound (with respect to the state and input
constraint sets) the size of the RCI set and its corresponding set of
control actions. Therefore, we employ the RCI LP problem as the key
ingredient in the following design procedure, for each coalition
$c \in \mc{C}$. The design starts with determining an RCI set for the
overall error, $e_c$, and the overall disturbance set $\mbb{W}_c$,
because the latter is known. The real aim is to determine an RCI
control law for the unplanned error, $\hat{e}_c$, and unplanned
disturbance set $\hat{\mbb{W}}_c$; however, the latter is not known
until the scaling constants $\alpha_c^x$ for each coalition have been
determined.

\begin{enumerate}
\item\label{step:first} The problem \eqref{eq:optRCI} associated with the dynamics
  $e_c^+ = A_{cc}e_c + B_{c}f_c + w_c$ and known constraint set
  $(\mbb{X}_c,\mbb{U}_c,\mbb{W}_c)$ is solved to yield
  $\gamma_{c,h} = (\mb{M}_{c,h},\eta_c,\theta_c,\delta_c)$, where
  $\eta_c$ and $\theta_c$ are scalings of $\mbb{X}_c$ and $\mbb{U}_c$
  such that $\mc{R}_{c,h} \subset \eta_c\mbb{X}_c$ and
  $\mu_c(\mc{R}_{c,h}) \subset \theta_c \mbb{U}_c$ respectively.
\item Given that, under the RCI control law $f_c = \mu_c(e_c)$,
  $e_c \in \mc{R}_{c,h} \subset \eta_c\mbb{X}_c$ and
  $f_c \in \mu(\mc{R}_{c,h}) \subset \theta_c \mbb{U}_c$, we select
  \begin{equation*}
  	\alpha_c^x = 1 - \eta_c;\quad 	\alpha_c^u = 1 - \theta_c,
  \end{equation*}
for the scaling factors in the main MPC problem. Then
$x_c = \bar{x}_c + e_c \in \alpha^x_c\mbb{X}_c \oplus \eta_c\mbb{X}_c
= \mbb{X}_c$, with a similar expression for $u_c$. The scaling
factor $\alpha_c^x$ is transmitted to neighbouring coalitions.
\item Given $\alpha_d^x$ for $d \in \mc{M}_c$, the
  set $\hat{\mbb{W}}_c = \bigoplus_{d \in \mc{M}_c} (1-\alpha^x_c)A_{cd} \mbb{X}_d$ is computed\footnote{Note that an outer-approximation to $\hat{\mbb{W}}_c$ is easily computed as $t_c\mbb{W}_c$ where $t_c = \left(\max_{d \in \mc{M}_c} (1-\alpha_d^x)\right)$. } and the RCI
  problem~\eqref{eq:optRCI}, now associated with $\hat{e}_c =
  A_{cc}\hat{e}_c + B_{c}\hat{f}_c + \hat{w}_c$ and
  $(\mbb{X}_c,\mbb{U}_c,\hat{\mbb{W}}_c)$, is re-solved for
  $\tilde{\gamma}_{(c,h)} =
  (\mb{M}_{c,h},\tilde{\eta}_c,\tilde{\theta}_c,\tilde{\delta}_c)$,
  yielding the scaling factors
  \begin{equation*}
  	 \xi_c^x =  \tilde{\eta}_c;\quad \xi_c^u =  \tilde{\theta}_c.	
  \end{equation*}
These scaling factors are such that ${\mc{R}}_{c,h} \subset
\xi^x_c\mbb{X}_c$ and $\mu_c({\mc{R}}_{c,h}) \subset
\xi^u_c\mbb{U}_c$; that is, the regions of the constraint sets that the
third-term robust control law occupies in response to the unplanned error and unplanned disturbance.
\item The selection of the constants $\beta_c^x$ and $\beta^u_c$ for
  the secondary MPC problem is made as
    \begin{equation*}
  	\beta_c^x = 1 - \alpha^x_c - \xi^x_c;\quad 	\beta_c^u = 1- \alpha^u_c - \xi^u_c.	
  \end{equation*}
Then $x_c = \bar{x}_c
+ \bar{e}_c + \hat{e}_c \in \alpha^x_c\mbb{X}_c\oplus \beta^x_c\mbb{X}_c \oplus \xi^x_c\mbb{X}_c = \mbb{X}_c$, as
required, with a similar expression for $u_c$.
\item The control law
  $\hat{f}_c = \tilde{\kappa}_c(\hat{e}_c) = \mu_c(\hat{e}_c)$ is
  computed from the matrices $\mb{M}_{c,h}$, using the minimal
  selection map procedure described in \citep{RM05}.
\end{enumerate}

\section{Proofs}

\subsection{Proof of Proposition~\ref{prop:feas}}

For part (i), because the nominal model is linear,
  $\alpha^x_c\mbb{X}_c$ and $\alpha^u_c\mbb{U}_c$ are PC-sets, and the
  terminal constraint is control invariant, the set $\bar{\mc{X}}_c^N$
  is compact, contains the origin and satisfies
  $\bar{\mc{X}}_c^N \supseteq \bar{\mc{X}}_c^{N-1} \supseteq \dots
  \supseteq \bar{\mc{X}}_c^0 = \{0\}$. Moreover, $\bar{\mc{X}}_c^N$ is
  positively invariant for
  $\bar{x}_c^+ = A_{cc}\bar{x}_c + B_{c}\bar{\kappa}_c(\bar{x}_c)$, which
  is sufficient to prove the claim. (For a detailed proof,
  see~\citep[Proposition 2.11]{RM_mpc_book}.) The same arguments applied to
  $\bar{\mc{E}}^H_c(\bar{\mb{w}})$ establish part (ii).

  For (iii), suppose that at time $k$, $\bar{x}_c \in \bar{\mc{X}}^N_c$,
  $\bar{e}_c \in \bar{\mc{E}}_c^H(\bar{\mb{w}}_c)$ with
  $\bar{\mb{w}}_c \in \bar{\mc{W}}^N_c$, and
  $\hat{e}_c \in {\mc{R}}_c$. Then
  ${x}_c \in \bar{\mc{X}}^N_c \oplus \bar{\mc{E}}_c^H(\bar{\mb{w}}_c)
  \oplus {\mc{R}}_c \subseteq \alpha^x_c \mbb{X}_c \oplus
  \beta^x_c \mbb{X}_c \oplus \xi^x_c \mbb{X}_c =
  (\alpha^x_c+\beta^x_c+\xi^x_c)\mbb{X}_c \subseteq \mbb{X}_c$. The
  applied control is
  $u_c = \bar{u}_c^0(0;\bar{x}_c) +
  \bar{f}_c^0(0;\bar{e}_c,\bar{\mb{w}}_c) + \mu_c(\hat{e}_c) \in
  \alpha^u_c\mbb{U}_c\oplus\beta^u_c\mbb{U}\oplus\xi^u_c\mbb{U}_c
  \subseteq \mbb{U}_c$. Then, because of parts (i) and (ii),
  ${x}_c^+ = A_{cc}x_c + B_{c}u_c + w_c \in \bar{\mc{X}}^N_c \oplus
  \bar{\mc{E}}_c^H(\bar{\mb{w}}_c^+) \oplus {\mc{R}}_c$. To
  complete the proof, however, we must consider the possibility that
  the disturbance sequence at the successor state is
  $\bar{\mb{w}}_c^0 \neq \bar{\mb{w}}_c^+$: in that case, if
  $\hat{\mbb{P}}_c(\bar{e}_c^+;\bar{\mb{w}}_c^0)$ is feasible then
  ${x}_c^+\in \bar{\mc{X}}^N_c \oplus \bar{\mc{E}}_c^H(\bar{\mb{w}}_c^0)
  \oplus {\mc{R}}_c$, which is still within $\mbb{X}_c$ by
  construction, and
  $u_c = \bar{u}_c^0(0;\bar{x}^+_c) +
  \bar{f}_c^0(0;\bar{e}^+_c,\bar{\mb{w}}^0_c) + \mu_c(\hat{e}^+_c)
  \subseteq \mbb{U}_c$. If
  $\hat{\mbb{P}}_c(\bar{e}_c^+;\bar{\mb{w}}_c^0)$ is not feasible,
  then $\hat{\mbb{P}}_c(\bar{e}_c^+;\bar{\mb{w}}_c^+)$ \emph{is}
  feasible (by the tail), and
  $u_c = \bar{u}_c^0(0;\bar{x}^+_c) +
  \bar{f}_c^0(0;\bar{e}^+_c,\bar{\mb{w}}^+_c) + \mu_c(\hat{e}^+_c)
  \subseteq \mbb{U}_c$. This establishes recursive feasibility of the
  algorithm.

  Finally, if, at time $0$, $\bar{x}_c = x_c \in \bar{\mc{X}}_c^N$ then
  $\bar{{e}}_c = 0$. Moreover, if $\bar{\mb{w}}_c = 0$,
  then---trivially---$\bar{{e}}_c \in \bar{\mc{E}}_c^H(0)$ and both
  the primary and secondary problems are feasible. By recursion,
  feasibility is retained at the next step, and the proof is complete.
  \qed

\subsection{Proof of Theorem~\ref{thm:stab}}

  By Proposition~\ref{prop:feas}, $\bar{x}_c \in \bar{\mc{X}}^N_c$
  implies
  $\bar{x}^+_c = A_{cc}\bar{x}_c + B_{c}\bar{\kappa}_c(\bar{x}_c) \in
  \bar{\mc{X}}^N_c$, with $\bar{V}_c^0(\bar{x}_c^+) \leq \bar{V}_c^0(\bar{x}_c) - \ell_c (\bar{x}_c,\bar{\kappa}_c(\bar{x}_c))$ where
    $\ell_c(x_c,u_c) \triangleq x^\top_c Q_c x_c + u^\top_c R_c
    u_c$. By Assumption~\ref{assump:pd}, there exists a constant
    $a_c > 0$ such that  \[\bar{V}_c^0(\bar{x}_c) \geq \ell_c(x_c,\bar{\kappa}_c(\bar{x}_c)) $ $\geq
    a_c |\bar{x}_c|^2\] for all $\bar{x}_c \in \bar{\mc{X}}_c^N$ and,
    moreover, Assumption~\ref{assump:controllability} ensures the
    existence of a constant $b_c > a_c > 0$ such that
    $\bar{V}_c^0(\bar{x}_c) \leq b_c |\bar{x}_c|^2$ over the same
    domain. Then $\bar{V}_c^0(\bar{x}_c^+) \leq \gamma_c \bar{V}_c^0(\bar{x}_c)$
    where $\gamma_c \triangleq (1 - a_c/b_c) \in (0,1)$. If
    $\bar{x}_c(0) \in\bar{\mc{X}}^N_c $ then
    $\bar{V}_c^0(\bar{x}_c(k)) \leq \gamma_c^k \bar{V}_c^0(\bar{x}_c(0))$ and 
    $\left| \bar{x}_c(k) \right| \leq d_c \delta_c^k \left|
      \bar{x}_c(0) \right|$ where
    $\delta_c \triangleq \sqrt{\gamma_c}$ and $d_c \triangleq \sqrt{b_c/a_c}$. This establishes exponential stability for the nominal system.

    Now consider the true trajectory $\{ x_c(k) \}_k$. We have
    $x_c(0) = \bar{x}_c(0) \in \bar{\mc{X}}_c^N$, so
    $|x_c(0)| = |\bar{x}_c(0)|$. Consider some
    $x_c = \bar{x}_c + e_c$; by Proposition~\ref{prop:feas},
    $e_c = \bar{e}_c + \hat{e}_c \in \bar{\mc{E}}_c^H\bigl(
    \bar{\mb{w}}_c \bigr) \oplus \hat{\mc{R}}_c$ implies
    $e_c^+ = \bar{e}_c^+ + \hat{e}_c^+ \in \bar{\mc{E}}_c^H\bigl(
    \bar{\mb{w}}_c^+ \bigr) \oplus \hat{\mc{R}}_c$, with
    \begin{equation*}
      V^H(\bar{e}^+_c,\bar{\mb{f}}_c^+; \bar{\mb{w}}_c^+) \leq \hat{V}^0_c(\bar{e}_c;\bar{\mb{w}}_c) - \ell_c (\bar{e}_c,\hat{\kappa}_c(\bar{e}_c;\bar{\mb{w}})).
    \end{equation*}
    At $\bar{e}_c^+$, the actual planned disturbance is
    $\tilde{\mb{w}}_c$, which may differ from $\bar{\mb{w}}^+_c$; this $\tilde{\mb{w}}_c$ is adopted if and only if $\bar{e}_c^+ \in \bar{\mc{E}}^H(\tilde{\mb{w}}_c)$ and $\hat{V}^0(\bar{e}_c^+; \tilde{\mb{w}}_c) \leq V^H(\bar{e}^+_c,\bar{\mb{f}}_c^+; \bar{\mb{w}}_c^+)$, implying that
    \begin{equation*}
      \hat{V}^0(\bar{e}_c^+; \tilde{\mb{w}}_c) \leq \hat{V}^0_c(\bar{e}_c;\bar{\mb{w}}_c) - \ell_c (\bar{e}_c,\hat{\kappa}_c(\bar{e}_c;\bar{\mb{w}}))
      \end{equation*}
      Thus, the sequence
      $\bigl\{ \hat{V}^0_c(\bar{e}_c(k); \bar{\mb{w}}_c(k) )
      \bigr\}_{k \in \mbb{I}_{\geq 0}}$ is bounded below by zero and
      monotonically decreasing, so the limit
      $\lim_{k \to \infty} \hat{V}^0_c(k) = 0$, which in turn implies
      $\bar{e}_c(k) \to 0$. Finally, we consider the dynamics of
      $\hat{e}_c$; since $\hat{e}_c = e_c - \bar{e}_c = x_c - \bar{x}_c - \bar{e}_c$ then
      \begin{equation*}
          \hat{e}_c^+ = A_{cc} \hat{e}_c + B_c\tilde{\kappa}_c(\hat{e}_c) + \sum_{d \in \mc{M}_c} A_{cd}\hat{e}_d.
        \end{equation*}
        Thus, $\hat{e}^+ = A\hat{e} + B\tilde{\kappa}(\hat{e})$, where
        $\tilde{\kappa}(\cdot)$ denotes the diagonal collection of
        $\tilde{\kappa}_c(\cdot)$, for which $\hat{e}(k) \to 0$ as
        $k \to \infty$, in view of
        Assumption~\ref{assump:decent}. Finally, since
        $x_c = \bar{x}_c + \bar{e}_c + \hat{e}_c$, and all three terms
        decay asympotically to zero, then $x_c (k) \to 0$ as
        $k \to \infty$.      
  \qed

  \subsection{Proof of Theorem~\ref{thm:FIP}}

   Consider a profile
    $\mc{C}(p) = \bigl(\mc{C}_{[1]}(p),\dots,\mc{C}_{[M]}(p)\bigr)$
    such that all $\mc{C}_{[i]}(p) \neq \mc{C}^e_{[i]}$ and that
    subsystem $i$ is to optimize. By definition, if $\Delta \geq 1$
    then there exists a
    $\mc{C}_{[i]}(p+1) \in \mbb{C}_\Delta(\mc{C}_{[i]}(p))$ such that
    \[J_i\bigl( \mc{C}_{[i]}(p+1), \mc{C}_{[-i]}(p)\bigr) < J_i\bigl(
    \mc{C}_{[i]}(p), \mc{C}_{[-i]}(p)\bigr).\]
    By repeated application,
    $\mf{C}$ is an improvement path. It must be finite because
    $\Pi_\mc{M}$ is a finite set. Thus, it terminates in a finite
    number of iterations to a Nash equilibrium, which is a minimum of
    the potential function.
    \qed

    \subsection{Proof of Theorem~\ref{thm:NE}}

    The potential function satisfies, for any $\mc{C}_{[i]} \in \mbb{C}_\Delta(\mc{C}_{[i]}^e)$
      \begin{equation*}
        \begin{split}
          \Delta \phi &\triangleq \phi( \mc{C}_{[1]}^e, \ldots,\mc{C}_{[i]},\ldots,\mc{C}_{[M]}^e) - \phi(\mc{C}_{[1]}^e,\ldots,\mc{C}_{[M]}^e)  \\
          &= \sum_{j\in\mc{M}_i} \left( w_{ij}(x_i,x_j)  \bigl| \delta_{ij}(\mc{C}_{[i]}) - \delta_{ij}(\mc{C}_{[j]}^e) \bigr|\right.
  \\
  &\quad \left. +  \rho \bigl( w_{ij}(x_i,x_j) + \epsilon \bigr)\sigma_{ij}\bigl[ \delta_{ij}(\mc{C}^e_{[i]}) - \delta_{ij}(\mc{C}_{[i]}) \bigr]\right),
  \end{split}
\end{equation*}
using the fact that $J_i^\ts{consensus} = 0$ at
$(\mc{C}_{[1]}^e,\dots,\mc{C}_{[M]}^e)$. Subsystem $i$ may either
refine $\mc{C}_{[i]}^e$ or coarsen it. If the latter, then
$\delta_{ij}(\mc{C}_{[i]}^e) \geq \delta_{ij}(\mc{C}_{[i]})$ for all
$j \in \mc{M}_i$ and all $i$, and, since the first term in
$\Delta \phi$ is always non-negative, the potential function can not
decrease. If the former, however, then second term in $\Delta \phi$
may be negative: we need to show that, if the provided bound holds and
$\epsilon$ is sufficiently small, then the overall difference in
potential is still positive.

Note that $w_{ij}(x_i,x_j) \bigl| \delta_{ij}(\mc{C}_{[i]}^e) -
  \delta_{ij}(\mc{C}_{[j]}^e) \bigr| = 0$ for all $j \in \mc{M}_i$,
  so, for all pairs $(i,j)$, either $w_{ij}(x_i,x_j) = 0$ or
  $\bigl| \delta_{ij}(\mc{C}_{[i]}^e) - \delta_{ij}(\mc{C}_{[j]}^e)
  \bigr| = 0$. Let $\tilde{\mc{M}}_i$ denote the subset of $\mc{M}_i$ for
  which $w_{ij}(x_i,x_j) > 0$. The potential difference is
  \begin{equation*}
    \begin{split}
  \Delta\phi &=   \sum_{j\in\tilde{\mc{M}}_i}  w_{ij}(x_i,x_j) \left( \bigl| \delta_{ij}(\mc{C}_{[i]}) - \delta_{ij}(\mc{C}_{[j]}^e) \bigr| \right. \\ &\quad \left. - \rho \sigma_{ij}\bigl[ \delta_{ij}(\mc{C}_{[i]}) - \delta_{ij}(\mc{C}^e_{[i]}) \bigr]\right)
  \\
  &\quad - \sum_{j \in \mc{M}_i} \rho \epsilon \sigma_{ij}\bigl[ \delta_{ij}(\mc{C}_{[i]}) - \delta_{ij}(\mc{C}^e_{[i]}) \bigr].
  \end{split}
\end{equation*}
For the first term, since $\bigl| \delta_{ij}(\mc{C}_{[i]}^e) - \delta_{ij}(\mc{C}_{[j]}^e)
\bigr| = 0$ then $\delta_{ij}(\mc{C}^e_{[i]}) = \delta_{ij}(\mc{C}^e_{[j]})$ for all $j \in \tilde{\mc{M}}_i$. It follows---also using the fact that $\delta_{ij}(\mc{C}_{[i]}^e) \leq \delta_{ij}(\mc{C}_{[i]})$ under refinement---that $\bigl| \delta_{ij}(\mc{C}_{[i]}) - \delta_{ij}(\mc{C}_{[j]}^e) \bigr| =\bigl[ \delta_{ij}(\mc{C}_{[i]}) - \delta_{ij}(\mc{C}^e_{[i]}) \bigr]$ and so this first term may be written
\begin{equation*}
\sum_{j\in\tilde{\mc{M}}_i} w_{ij}(x_i,x_j)(1 - \rho\sigma_{ij})\bigl[ \delta_{ij}(\mc{C}_{[i]}) - \delta_{ij}(\mc{C}^e_{[i]}) \bigr],
  \end{equation*}
  which is non-negative if $\rho\sigma_{ij} < 1$ for all
  $j \in \mc{M}_i$. We now consider the subtraction of the second term
  from this one. Suppose that $\rho \sigma_{ij} \leq \gamma < 1$ for
  all $i$ and $j$. Then
  \begin{equation*}
\begin{split}
    \Delta \phi &\geq (1-\gamma) \sum_{j \in \tilde{\mc{M}}_i} w_{ij}(x_i,x_j) \bigl[ \delta_{ij}(\mc{C}_{[i]}) - \delta_{ij}(\mc{C}^e_{[i]}) \bigr] \\
    & \quad - \gamma \epsilon \sum_{j \in \mc{M}_i}   \bigl[ \delta_{ij}(\mc{C}_{[i]}) - \delta_{ij}(\mc{C}^e_{[i]}) \bigr].
    \end{split}
    \end{equation*}
    It follows that $\Delta \phi \geq 0$ if, for all $i \in \mc{M}$,
    \begin{equation*}
      \epsilon \leq \frac{1 - \gamma}{\gamma} \frac{\sum_{j \in \tilde{\mc{M}}_i} w_{ij}(x_i,x_j) \bigl[ \delta_{ij}(\mc{C}_{[i]}) - \delta_{ij}(\mc{C}^e_{[i]}) \bigr]}{\sum_{j \in \mc{M}_i}   \bigl[ \delta_{ij}(\mc{C}_{[i]}) - \delta_{ij}(\mc{C}^e_{[i]}) \bigr]}.
    \end{equation*}
    \qed

\subsection{Proof of Theorem~\ref{thm:nest}}

The first part of the proof is to show that if $\mc{C} \succeq \mc{D}$, then
  \begin{equation*}
  \mbb{W}_\mc{C} = \prod_{c \in \mc{C}} \bigoplus_{d \in \mc{M}_c} A_{cd} \mbb{X}_d \subseteq \prod_{c \in \mc{D}} \bigoplus_{d \in \mc{M}_c} A_{cd} \mbb{X}_d = \mbb{W}_\mc{D}.
\end{equation*}
Consider an arbitrary partition $\mc{C} = \{1,\dots,C \} \in \Pi_\mc{M}$, and a refinement $\mc{D} = \{1,\dots,C-1,C^\prime,C^\prime+1\}\}$; that is, coalition $C$ in the first partition is split into two coalitions, containing subsystems $C^\prime$ and ${C^\prime+1}$ such that ${C^\prime} \cup (C^\prime+1) = {C}$. If the set of neighbours for coalition $C$ is $\mc{M}_C = \left\{ d \in \{1,\dots,C-1\} : A_{Cd} \neq 0 \right\} $ then the sets of neighbours for the new coalitions are
\begin{align*}
  \mc{M}_{C^\prime} &= \left\{ d \in \{1,\dots,C-1,C^\prime+1\}  : A_{C^\prime d} \neq 0 \right\}\\
 \mc{M}_{C^\prime+1} &= \left\{ d \in \{1,\dots,C-1,C^\prime\}  : A_{(C^\prime+1)d} \neq 0 \right\}
  \end{align*}
  such that $\mc{M}_{C^\prime} \setminus \mc{M}_C = \{ C^\prime + 1\}$ and $\mc{M}_{C^\prime+1} \setminus \mc{M}_C = \{ C^\prime \}$. Then
  \begin{equation*}
\mbb{W}_{C^\prime} = \bigoplus_{d \in \mc{M}_{C^\prime}} A_{C^\prime d} \mbb{X}_d 
      =  \bigoplus_{j \in  \mc{H}_{C^\prime}} \prod_{i \in {C^\prime}} A_{ij} \mbb{X}_j  
%    \begin{split}
%       =\bigoplus_{d \in \mc{M}_{C^\prime}} A_{C^\prime d} \prod_{j \in d} \mbb{X}_j\\
%%      &= \bigoplus_{d \in \mc{M}_{C^\prime}} \prod_{i \in {C^\prime}} \bigoplus_{j \in d}  A_{ij} \mbb{X}_j\\
%%      &= \prod_{i \in {C^\prime}} \bigoplus_{j \in \bigcup\limits_{d \in \mc{M}_{C^\prime}}d } A_{ij} \mbb{X}_j \\
%      &=  \bigoplus_{j \in  \mc{H}_{C^\prime}} \prod_{i \in {C^\prime}} A_{ij} \mbb{X}_j
%      \end{split}
  \end{equation*}
  where $\mc{H}_{C^\prime} =\bigcup_{d \in \mc{M}_{C}'} d $, and with a similar expression for $\mbb{W}_{C^\prime+1}$.
  Note also that $\mbb{W}_{C} = \bigoplus\limits_{j \in \mc{H}_i} \prod_{i \in {C}}  A_{ij} \mbb{X}_j$ with $\mc{H}_C =\bigcup_{d \in \mc{M}_{C}} d $ and then
  \begin{equation*}
    \mbb{W}_{C^\prime} \times \mbb{W}_{C^\prime + 1} = \bigoplus_{j \in \tilde{\mc{H}}_i} \prod_{i \in {C}}  A_{ij} \mbb{X}_j \supseteq \mbb{W}_{C}
        \end{equation*}
        where $\tilde{\mc{H}}_C = \bigcup_{d \in ( \mc{M}_{C^\prime} \cup \mc{M}_{C^\prime+1} )} d $ and the latter inclusion follows from the fact that $\mc{M}_{C^\prime} \cup \mc{M}_{C^\prime+1} \supseteq \mc{M}_C$.  Since $\mbb{W}_\mc{C} = \prod_{c \in \mc{C}} \mbb{W}_c$, it follows that $\mbb{W}_{\mc{C}} \subseteq \mbb{W}_{\mc{D}}$; since $\mc{C}$ and $\mc{D}$ were arbitrary, the result holds in general.

        The second and final part is to show that $\mbb{W}_{\mc{C}} \subseteq \mbb{W}_{\mc{D}}$ implies $\mc{R}_{\mc{C}} \subseteq \mc{R}_{\mc{D}}$. Consider again the refinement $\mc{D}$ and suppose the RCI sets for the $C^\prime+1$ coalitions are $\mc{R}_1,\dots,\mc{R}_{C-1},\mc{R}_{C^\prime},\mc{R}_{C^\prime+1}$. The product of the latter two coalitional sets is $\mc{R}_{C^\prime} \times \mc{R}_{C^\prime+1}$, associated with the disturbance set $\mbb{W}_{C^\prime} \times \mbb{W}_{C^\prime + 1}$. Since $\mbb{W}_{C^\prime} \times \mbb{W}_{C^\prime + 1} \supseteq \mbb{W}_{C}$, then there exist two scalars $0 < a < b \leq 1$ such that
        \begin{equation*}
          a(\mbb{W}_{C^\prime} \times \mbb{W}_{C^\prime + 1}) \subseteq \mbb{W}_{C} \subseteq b(\mbb{W}_{C^\prime} \times \mbb{W}_{C^\prime + 1}).
          \end{equation*}
          Then consider the set
          \begin{equation*}
           \bigoplus_{l=0}^{h-1} D_l(\mb{M}_h) b (\mbb{W}_{C^\prime} \times \mbb{W}_{C^\prime + 1}) = b (\mc{R}_{C^\prime} \times \mc{R}_{C^\prime+1}).
          \end{equation*}
          Since $b(\mc{R}_{C^\prime} \times \mc{R}_{C^\prime+1}) \subseteq (\mc{R}_{C^\prime} \times \mc{R}_{C^\prime+1})$ is RCI for $b(\mbb{W}_{C^\prime} \times \mbb{W}_{C^\prime + 1})$, it must be RCI for any subset of $b(\mbb{W}_{C^\prime} \times \mbb{W}_{C^\prime + 1})$, including $\mbb{W}_{C}$. Thus, $b(\mc{R}_{C^\prime} \times \mc{R}_{C^\prime+1})$ outer-bounds $\mc{R}_C$.\qed

\subsection{Proof of Theorems~\ref{thm:final} and~\ref{thm:conv}}

Let ${x}(0)\in \bar{\mc{X}}^N_{\mc{C}(0)}$, where the initial
partition is $\mc{C}(0)$. Since
$\bar{\mc{X}}_{\mc{C}(0)}^N\subset\Rset^n$ is compact, there exists an
$r>0$ such that
${x}(0)\in \mc{B}_r \triangleq \{ x : \left| x \right| \leq r \}
$. Take
$\zeta_{\mc{C}(0)} = \min_{\bar{x}} \left\{ \bar{V}^0_{\mc{C}(0)}(
  \bar{x} ) : |\bar{x}| =r \right\}$, where
$\bar{V}^0_{\mc{C}(0)}( \bar{x} )$ denotes the collective value
functions of the primary MPC controllers, \ie~the sum of
$\bar{V}^0_c(\bar{x}_c)$ over $c \in \mc{C}(0)$. The associated
sub-level set is $S(\zeta_{\mc{C}(0)}) =\left\{ \bar{x} : \bar{V}^0_{\mc{C}(0)}( \bar{x} )  \leq \zeta_{\mc{C}(0)} \right\} \subseteq\mc{B}_r$. In line with the hypothesis on the switching signal, the system
remains in the partition $\mc{C}(0)$ for a number of time steps; call
this number $k_s$. By Theorem~\ref{thm:stab}, the value function at
time $k = k_s$ is bounded as $  V_{\mc{C}(0)}\left(\bar{x}(k_s)\right) \leq \gamma_{\mc{C}(0)}^{k_s} \zeta_{\mc{C}(0)}$ where $\gamma_{\mc{C}(0)} = \max_{c \in \mc{C}(0)} \gamma_c$, and
$\gamma_c \in (0,1)$ is the decay constant for the primary controller
for coalition $c$ from Theorem~\ref{thm:stab}.

Suppose that, at $k=k_s$, the strongly feasible partition
$\mc{C}(k_s) \in \Pi_\mc{M}$ is selected, such that
$x(k_s) \in \bar{\mc{X}}^N_{\mc{C}(k_s)}$. The value functions
$\{\bar{V}^0_c(\cdot)\}_{c \in \mc{C}(k_s)}$ are bounded on the
individual sets $\bar{\mc{X}}^N_c$ such that
$\theta_{\mc{C}} \triangleq \sup\left\{ \bar{V}^0_{\mc{C}(k)} (\bar{x}) : \bar{x} \in \bar{\mc{X}}^N_\mc{C} \right\}$ is attained as the
  maximum, and is finite, for any $\mc{C} \in \Pi_\mc{M}$; in
  addition, define
  $\mu_{\mc{C}} =
  \zeta_{\mc{C}}^{-1}(\theta_{\mc{C}}+\epsilon_{\mc{C}})$ for an
  $\epsilon_{\mc{C}}>0$, and
  $\mu = \max \left\{ \mu_{\mc{C}} : \mc{C} \in \Pi_\mc{M}
  \right\}$. The value function for $\mc{C}(k_s)$ satisfies
  $V_{\mc{C}(k_s)}\left(\bar{x}(k_s)\right) \leq \theta_{\mc{C}(k_s)}
  < \theta_{\mc{C}(k_s)} + \epsilon_{\mc{C}(k_s)} \leq \mu
  \zeta_{\mc{C}(0)} = \mu V_{\mc{C}(0)}
  \left(\bar{x}(0)\right)$. Thus, the
  conditions of Lemma~\ref{lem:stable_switching} are met, and the system is
  locally stable. The final part of the proof establishes attractivity of the origin,
  as sufficient to prove Theorem~\ref{thm:conv}. Owing to the previous
  result, the switched value function remains bounded, and the system
  remains feasible by hypothesis, for all $k \in [0, k_f]$. For
  $k \geq k_f$, it is assumed that $\sigma(k) = \bar{\mc{C}}$

 \qed

\end{document}